\documentclass[12pt,preprint]{aastex}

\shorttitle{PHOEBE - PHysics Of Eclipsing BinariEs}
\shortauthors{Pr\v sa, Zwitter}

\newcommand {\phoebe}{{\tt PHOEBE }}
\newcommand {\phoebents}{{\tt PHOEBE}} 
\newcommand {\kms}{\mathrm{km~s}^{-1}}

\newcommand {\RSun}{\mathrm R_\odot}
\newcommand {\MSun}{\mathrm M_\odot}
\newcommand {\LSun}{\mathrm L_\odot}

\newcommand {\Teff}{T_\mathrm{eff}}
\newcommand {\logg}{\log\,\,(g/g_0)}
\newcommand {\vrot}{v_\mathrm{rot}}
\newcommand  {\hla}{L_1^i}
\newcommand  {\vga}{v_\gamma}

\begin{document}

\title{A Computational Guide to Physics of Eclipsing Binaries. \\ Paper I. Demonstrations and Perspectives}

\author{A. Pr\v sa}
\affil{University of Ljubljana, Department of Physics, Jadranska 19, 1000 Ljubljana, Slovenia}
\email{andrej.prsa@fmf.uni-lj.si}

\and

\author{T. Zwitter}
\affil{University of Ljubljana, Department of Physics, Jadranska 19, 1000 Ljubljana, Slovenia}
\email{tomaz.zwitter@fmf.uni-lj.si}

\begin{abstract}
\phoebe ({\tt PH}ysics {\tt O}f {\tt E}clipsing {\tt B}inari{\tt E}s) is a modeling package for eclipsing binary stars, built on top of the widely used {\sf WD} program \citep{wilson1971}. This introductory paper overviews most important scientific extensions (incorporating observational spectra of eclipsing binaries into the solution-seeking process, extracting individual temperatures from observed color indices, main-sequence constraining and proper treatment of the reddening), numerical innovations (suggested improvements to {\sf WD}'s Differential Corrections method, the new Nelder \& Mead's downhill Simplex method) and technical aspects (back-end scripter structure, graphical user interface). While \phoebe retains 100\% {\sf WD} compatibility, its add-ons are a powerful way to enhance {\sf WD} by encompassing even more physics and solution reliability. The operability of all these extensions is demonstrated on a synthetic main-sequence test binary; applications to real data will be published in follow-up papers. \phoebe is released under the GNU General Public License, which guarranties it to be free, open to anyone interested to join in on future development.
\end{abstract}


\section{Introduction}

With the ever-growing computer power, numerical models built to analyze acquired eclipsing binary data are gaining both on accuracy and complexity. The motivation is clear: due to their unique geometrical and kinematic properties, eclipsing binaries (EBs) give full physical insight into their structure, distance and evolution stage of their coeval components. In the last 40 years the EB field was overwhelmed by many approaches to solution seeking; \citet{kallrath1999} give an overview of most important ones. The widely used {\sf WD} code \citep{wilson1971} underwent many expansions, improvements and fine-tuning (\cite{wilson1976, wilson1979, wilson1990, milone1992, kallrath1998, vanhamme2003} and many others), which firmly established it as the most prominent software available for EBs.

So why would one build yet another modeling program? The answer is simple: one \emph{would not}. Tackling same old problems all over again does not make sense; rather, one builds on basis of what has already been done. This is what our effort is all about: to create a modeling package built on top of the Wilson--Devinney code, introducing new enhancements to where {\sf WD} was deficient, while still pertaining 100\% {\sf WD} compatibility. Enhancements include new physics (proper handling of color indices and therefore temperatures in absolute units, interstellar reddening effects), existing minimization scheme add-ons (stability and convergence improvements) and new minimization schemes aiming to fully automate first steps of solution-seeking (an issue of utmost importance for ambitious space scanning missions like Gaia \citep{perryman2001}). We discuss main characteristics of this new package called \phoebents: \verb|PH|ysics \verb|O|f \verb|E|clipsing \verb|B|inari\verb|E|s.

This paper introduces the formalism \phoebe is built on and demonstrates its capabilities on synthetic binary data that are described in Section \ref{teststar}. Sections \ref{inverse} and \ref{physics} give a detailed overview of computational and physical extensions. Section \ref{conclusion} discusses further work to be done and explains future vision of this project. Technical details on \phoebe availability and license, back-end logic and structure, and front-end interface are given in the Appendix.

\section{Building a test binary star} \label{teststar}

To demonstrate innovations \phoebe brings to the EB field, a synthetic binary model is created. Testing the methods against a synthetic model may seem artificial, but the obvious advantage of knowing the right solutions is the only true way of both qualitative and quantitative assessment. Some preliminary results of using \phoebe on true observations were already presented by \citet{prsa2003}. Full-fledged demonstration of \phoebe capabilities both for individual stars and large data-sets will be published in series of follow-ups to this paper shortly.

\begin{figure}
\includegraphics[width=16cm,height=9.6cm]{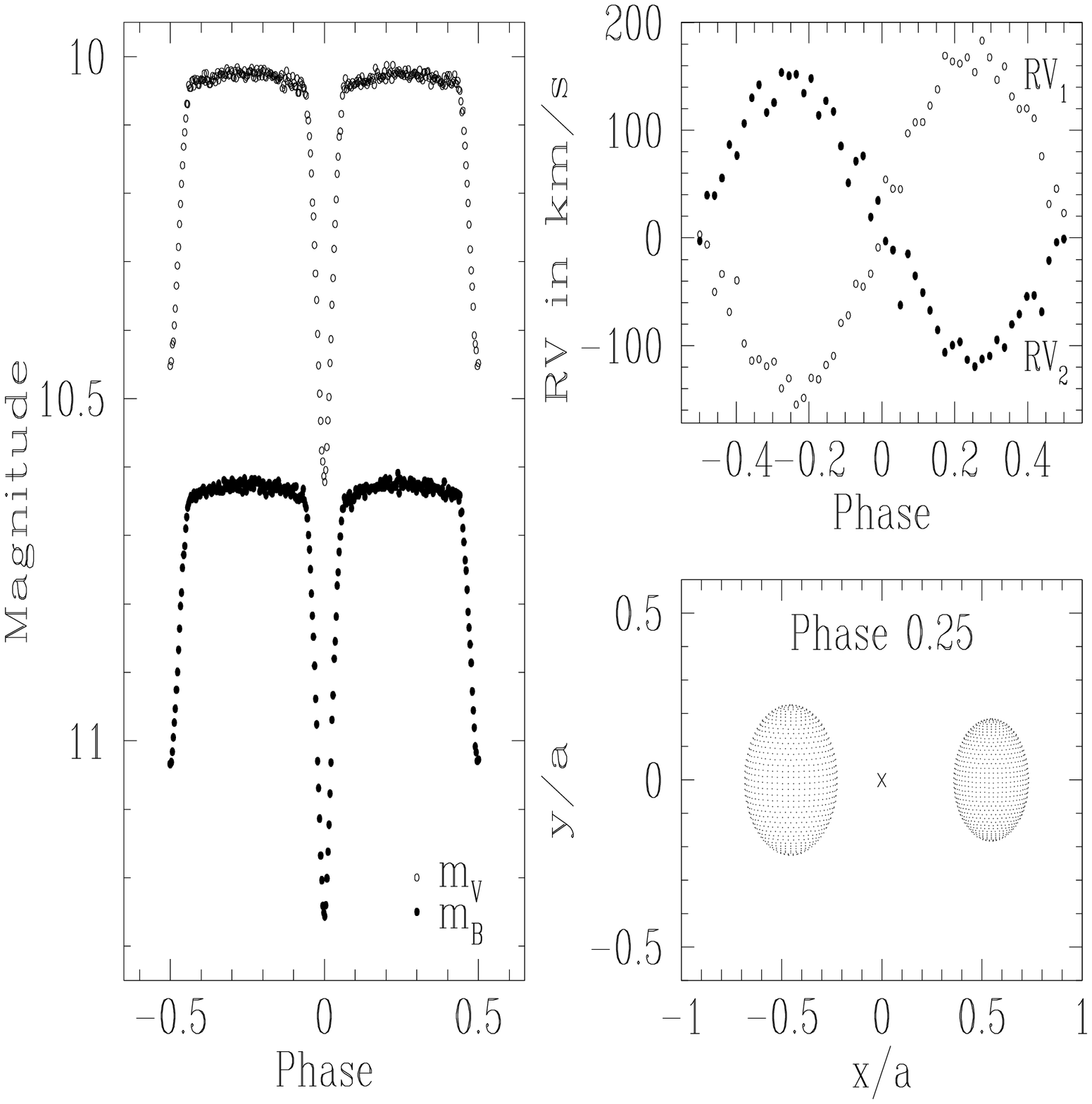} \\
\caption{F8~V--G1~V test star data. Light curves are computed for Johnson B (filled dots) and V (empty dots) passbands in 300 phase points with $\sigma_\mathrm{V} = 0.015$ (left panel). RV curves are computed in 50 phase points with $\sigma_\mathrm{RV} = 15\,\kms$ (upper right panel); eclipse proximity effects are turned off. Star plot is computed at quarter phase, cross denotes the center-of-mass (lower right panel).} \label{synthetic_lc_rv_starplot}
\end{figure}

Our synthetic binary consists of two main-sequence F8~V--G1~V components with their most important orbital and physical parameters listed in Table \ref{synthetic_binary}. It is a partially eclipsing detached binary with only slight shape distortion of both components ($R_{1,\mathrm{pole}}/R_{1,\mathrm{point}} = 0.974$, $R_{2,\mathrm{pole}}/R_{2,\mathrm{point}} = 0.979$). Light curves are generated for Johnson B and V passbands in 300 phase points with Poissonian scatter ranging from $\sigma_{\mathrm{V}} = 0.005$ to $\sigma_{\mathrm{V}} = 0.025$ at quarter-phase magnitude $m_V = 10.0$. Radial velocity (RV) curves are generated in 50 phase points with Gaussian scatters ranging from $\sigma_{\mathrm{RV}} = 1\,\kms$ to $\sigma_{\mathrm{RV}} = 25\,\kms$. Light curves in B and V with $\sigma_\mathrm{V} = 0.015$ and both RV curves with $\sigma_{\mathrm{RV}} = 15\kms$ are depicted in Fig.~\ref{synthetic_lc_rv_starplot}.

\begin{deluxetable}{lrrr}
\tablecaption{\label{synthetic_binary} Physical parameters of the F8~V--G1~V test binary star. Spectral type -- temperature relation taken from \cite{lang1992}.}
\tablehead
  {
  \colhead{Parameter [units]} &                & \colhead{Binary} & \\
                              & \colhead{F8~V} & & \colhead{G1~V}
  }
\startdata
$P_0$ [days]                             &        &  1.000 &        \\
$a\,\,\,[\RSun]$                         &        &  5.524 &        \\
$q=m_2/m_1$                              &        &  0.831 &        \\
$i\,\,\,[{}^\circ]$                      &        & 85.000 &        \\
$v_\gamma\,\,\,[\kms]$                   &        & 15.000 &        \\
$T_\mathrm{eff}\,\,\,[\mathrm K]$        &   6200 &        &   5860 \\
$L\,\,\,[\LSun]$                         &  2.100 &        &  1.100 \\
$M\,\,\,[\MSun]$                         &  1.239 &        &  1.030 \\
$R\,\,\,[\RSun]$                         &  1.260 &        &  1.020 \\
$\Omega\,\,\,[-]\tablenotemark{a}$       &  5.244 &        &  5.599 \\
$\logg\,\,\,[-]\tablenotemark{b}$        &  4.33  &        &  4.43  \\
$x_B\,\,\,[-]\tablenotemark{c}$          &  0.818 &        &  0.833 \\
$y_B\,\,\,[-]\tablenotemark{c}$          &  0.203 &        &  0.158 \\
$x_V\,\,\,[-]\tablenotemark{c}$          &  0.730 &        &  0.753 \\
$y_V\,\,\,[-]\tablenotemark{c}$          &  0.264 &        &  0.242 \\
\enddata
\tablenotetext{a}{Unitless effective potentials defined by \cite{wilson1979}.}
\tablenotetext{b}{$g_0 = 1 \mathrm{cm}\,\mathrm{s}^{-2}$ is introduced so that the logarithm acts on a dimensionless variable.}
\tablenotetext{c}{$\mathrm{L}$inear (x) and non-linear (y) coefficients of the logarithmic limb darkening law for Johnson B and V passbands, taken from \citet{vanhamme1993}.}
\end{deluxetable}

This model binary will be used for demonstrating all \phoebents's capabilities that are novel to the field of EBs.

\section{Solving the inverse problem for eclipsing binaries} \label{inverse}

The underlying {\sf WD} code is composed of two parts: the \verb|LC| program for computing light and RV curves and the \verb|DC| program for solving the inverse problem \citep{wilson1993}. \phoebe introduces several optimizations to the \verb|DC| method and adds to generality by implementing a new minimization method: Nelder \& Mead's downhill Simplex.

\subsection{Suggested optimizations to {\sf WD} solving method}

{\sf WD}'s \verb|DC| code, as the name suggests, uses Differential Corrections (DC) method complemented by the Levenberg-Marquardt algorithm to solve the inverse problem \citep{wilson1993}. It is especially suited for EBs and is one of the fastest codes around. In cases when the method does not converge, the Method of Multiple Subsets (MMS) may be used to relax the system to the nearest minimum \citep{wilson1976a}.

A \verb|DC| program reads in a user-supplied input file consisting of {\bf a)} a set of initial parameters that define physical and geometrical properties, {\bf b)} observational data and {\bf c)} switches that define the way a minimization algorithm is run (refer to the booklet by \citet{wilson2003} accompanying {\sf WD} code for details on \verb|DC| input files). Within one iteration, the values of parameters set for adjustment are improved and returned for user inspection. In case of convergence, the user manually resubmits the new parameter set to the next iteration. The measure of the quality of the fit (the cost function) is the sum of squares of weighted $O-C$ residuals.

{\sf WD}'s list of more than 30 adjustable parameters includes passband luminosities $\hla$ for $i$ light curves (with their {\sf WD} name {\sf HLA}), that have a unique property of \emph{linearly} scaling the level of light curves. \verb|DC| (or any other minimization algorithm) fits these luminosities the same way it fits all other physical parameters: softly. This means that within one iteration, the values of $\hla$ are \emph{not} fully adjusted, only improved. Since $\hla$'s determine vertical offset of light curves, this raises two specific problems: {\bf 1)} The soft change of $\hla$ in every iteration step causes \emph{artificial} changes of other physical parameters: rather than fitting the shape of the data curve, other parameters fit the \emph{discrepancy} between the model and the data, induced by the softness of $\hla$ fit. It is like driving a very old car on a very bumpy road - each bump on the road causes wobbling of the whole car with slow attenuation. {\bf 2)} Changes of adjusted parameters calculated by {\tt DC} will properly contribute to the cost function \emph{only} if the model is aligned with the data: the average $O-C$ value must be approximately 0. This alignment is governed by $\hla$ for light curves. If this alignment is not computed correctly, the cost function is \emph{misleading} DC instead of aiding it. This causes under-estimation of formal errors due to $\hla$ softness error propagation and even convergence problems.

\begin{figure}[htb]
\begin{center}
\includegraphics[width=16cm, height=5.33333cm]{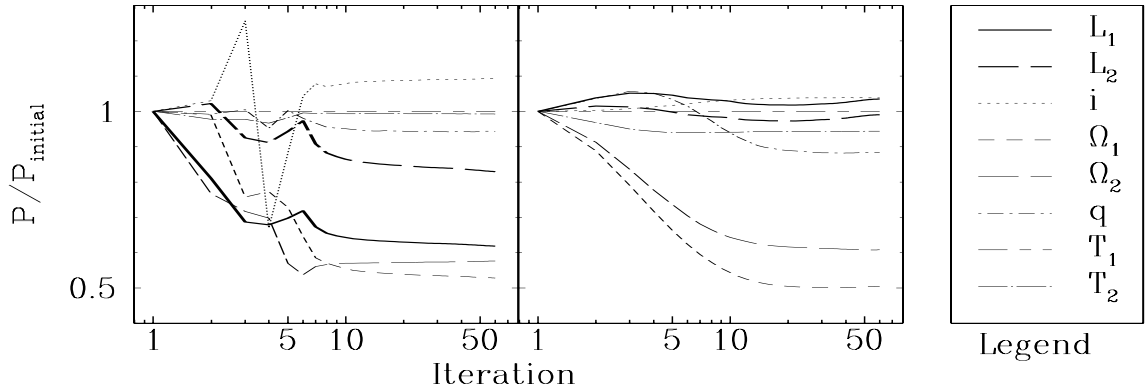} \\
\caption{\label{hla_stiffness} Soft vs.~stiff curve leveling. Iteration sequence for 7 physical parameters of our test binary for soft $\hla$ scheme (left) and stiff $\hla$ scheme (right). The x-axis is given in log-scale to amplify the part where the impact of stiffening is largest. Unity on $y$-axis corresponds to parameter's initial value. $L_1$ and $L_2$ are passband luminosities in B and V filter, respectively, $i$ is the system inclination, $\Omega_1$ and $\Omega_2$ are gravity potentials, $q$ is the mass ratio and $T_1$ and $T_2$ are surface temperatures. Temperature $T_1$ is kept constant throughout the fit, simulating the usual practice of determining one temperature and fitting the other.}
\end{center}
\end{figure}

\phoebe solves this problem by supplying an option to \emph{compute} $\hla$'s instead of minimizing them, thus increasing their stiffness with respect to other parameters. The alignment is calculated so that the average $O-C$ value is \emph{exactly} 0. The time cost of this computation is not only negligible, it actually speeds up the overall algorithm, since the dimension of the parameter subspace submitted to {\tt DC} is reduced. Fig.~\ref{hla_stiffness} demonstrates the iteration sequence with the original method (left) and the proposed method (right) for a case of 7 simultaneously fitted parameters displaced by at most 50\% from their true value. In the latter case parameters converge quickly and in a smooth fashion. Similar simulations that test convergence behavior in cases when both temperatures are fitted or when other individual parameters are kept constant have also been performed; they accord or even amplify the conclusion of Fig.~\ref{hla_stiffness} and their results are thus omitted on account of brevity. Note however that stiffening $\hla$'s does not guarantee convergence to the global minimum, it only solves the inverse problem more efficiently. It should also be stressed that calculating $\hla$'s instead of fitting them might not always affect convergence as noticeably, particularly in cases where relative corrections of parameter values are small.

By calculating $\hla$'s instead of fitting them, the $\chi^2$ criterion is not used and the corresponding formal errors of $\hla$'s are not calculated. To obtain them, one would simply revert from calculating to fitting $\hla$'s at the very end of the minimization process and submit them to the final iteration of the DC.

\paragraph{Systemic velocity $\vga$.}

The levels of RV curves are determined by the systemic velocity $\vga$: changing it vertically shifts those curves. Although $\vga$ is not as correlated with other parameters as is the case for $\hla$'s and the problem is thus not as severe, alignment between the model and the data is still crucial. \phoebe allows $\vga$ calculation following the same logic as before for $\hla$'s -- by demanding that the average $O-C$ value is exactly 0.

\paragraph{Limb darkening coefficients.}

The native {\sf WD} code supports linear, logarithmic and square root limb darkening (LD) laws. Their coefficients primarily depend on the given passband, effective temperature, gravity acceleration $\logg$ and metallicity [M/H]. {\sf WD} does not constrain the choice of these coefficients, so people have traditionally used LD tables computed by, e.g., \cite{vanhamme1993} or \cite{claret2000}.

Following a similar argument to the one mentioned before for $\hla$'s and $v_\gamma$, \phoebe implements an optional dynamical LD computation. After each iteration that induces changes to any of the $\Teff$, $\logg$, [M/H] or related parameters, the LD coefficients need to be modified accordingly. \phoebe uses \cite{vanhamme1993} tables for this purpose, dynamically reading out tabulated values and linearly interpolating to obtain proper values automatically. The implications are not as severe as for the $\hla$'s and $v_\gamma$ because LD contributions are orders of magnitude smaller and insensitive to small changes in the above mentioned parameters.

\subsection{New minimization algorithms} \label{minimization}

The main driving force of any binary minimization algorithm is its ability to solve the inverse problem as accurately and as quickly as possible. {\sf WD}'s DC algorithm is very fast and works well if the discrepancy between the observed and computed curves is relatively small, but it can diverge or give physically implausible results if the discrepancy is large. While this deficiency is usually not a severe problem when analysing individual EBs (one can always obtain a reasonable set of starting parameters by calculating a few initial light and RV curves), its impact when dealing with huge data-sets (such as hundreds of thousands of light curves that will be obtained by Gaia) may be a blocker. To overcome this, and to assist in initial steps of solution-seeking, a complementary minimization scheme to DC is proposed.

\paragraph{Nelder \& Mead's downhill Simplex.}

Two main deficiencies of DC are especially striking. {\bf 1)} The main source of divergence and the loss of accuracy in DC is the computation of numerical derivatives of the cost function with respect to parameters set for adjustment. {\bf 2)} Once DC converges, there is no ready way of telling whether the minimum is local or global; the method cannot escape. The latter problem affects most minimization algorithms that have been applied to EBs.

To circumvent these two problems, \phoebe implements Nelder \& Mead's downhill Simplex\footnote{Nelder \& Mead's downhill Simplex should not be confused with linear or non-linear programming algorithms, which are also referred to as Simplex methods \citep[e.g.][]{nr1992}.} method \citep{nelder1965}, hereafter NMS. Since NMS does not compute derivatives but relies only on function evaluations, it cannot diverge. The basic form of NMS applied to a {\sf WD} implementation was first proposed by \cite{kallrath1987}. \phoebe goes a step further and adapts the method specifically to EBs. First tests of \phoebents's NMS implementation on photometric data that are expected to be obtained by Gaia \citep{prsa2005} are very promising.

NMS acts in $n$-dimensional parameter hyperspace. It constructs $n$ vectors $\mathbf p_i$ from the vector of initial parameter values $\mathbf x$ and the vector of step-sizes $\mathbf s$ as follows:
\begin{equation}
\mathbf p_i = (x_0, x_1, \dots, x_{i-1}, x_i + s_i, x_{i+1}, \dots, x_n)
\end{equation}

\begin{figure}
\begin{center}
\includegraphics[width=12cm,height=6cm]{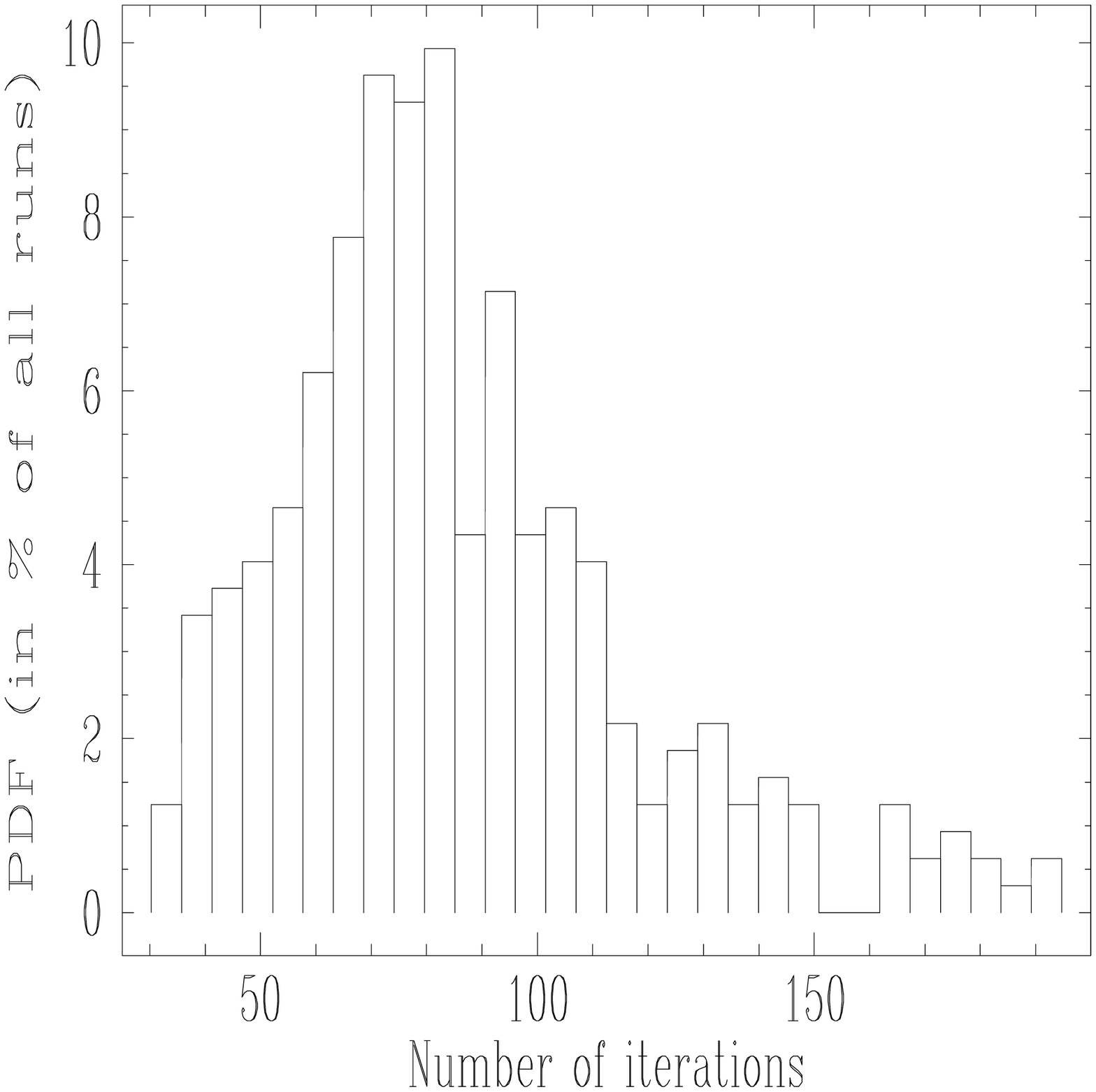}
\caption{Histogram of the number of iterations required for NMS convergence within $10^{-3}$ fractional accuracy. The Probability distribution function (PDF) exhibits a maximum at $\sim$75 iterations. Due to extremely fast convergence in the first few steps, the number of iterations is in practice insensitive to the selection of the initial starting point in parameter hyperspace; the required number of iterations is dominated by convergence behavior in the "Minima valley".} \label{nms_poisson}
\end{center}
\end{figure}

These vectors form $(n+1)$ vertices of an $n$-dimensional simplex. During each iteration the algorithm tries to improve parameter vectors $\mathbf p_i$ by modifying the vertex with the highest function value by simple geometrical transformations: reflection, reflection followed by expansion, contraction and multiple contraction \citep{galassi2003}. Using these transformations, the simplex moves through parameter space towards the closest minimum, where it contracts itself. Fig.~\ref{nms_poisson} shows the number of iterations required for the NMS to converge from different starting points within $10^{-3}$ fractional accuracy. The data are those of our test binary given in Fig.~\ref{synthetic_lc_rv_starplot}.

This basic form of NMS is unconstrained, which means that parameters may assume any value regardless of their physical feasibility. The NMS implemented by \phoebe optionally enables semi-constrained or fully constrained minimization by imposing limits to several or all adjusted parameter values. Additionally, heuristic scan, parameter kicking and conditional constraining enable NMS to efficiently escape from local minima.

\subsection{Heuristic Scan}

EB minimization algorithms, including even NMS with its property of guarranteed convergence, can be stuck in a local minimum, particularly since parameter hyperspace in vicinity of the global minimum is typically very flat, with lots of local minima. In addition, global minimum may be shadowed by data noise and degeneracy.

Heuristic scan is an enhancement method to any minimization algorithm (DC, NMS, \dots) that selects a set of starting points in parameter hyperspace and starts the minimization from each such point. The user defines how starting points are selected -- they may be gridded, stochastically dispersed, distributed according to some probability distribution function (PDF) etc. The algorithm then sorts all solutions by the cost function (the $\chi^2$, for example) and weights the obtained parameter values accordingly: heuristic runs with smallest values of the cost function correspond to the deepest minima and should thus be most weighted -- they are most suitable candidates for the global minimum.

\begin{figure}
\begin{center}
\includegraphics[width=16cm,height=5.33333cm]{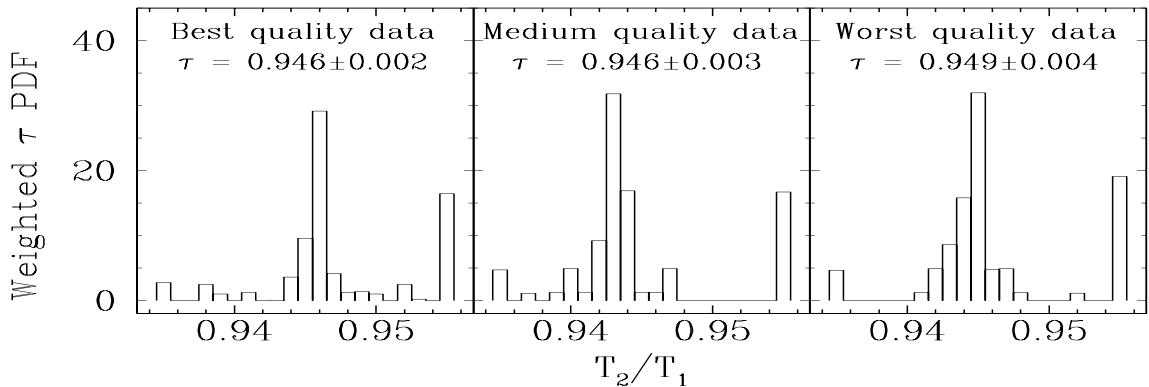}
\caption{Temperature ratio histogram obtained as a result of heuristic scan. Plots show $\tau = T_2/T_1$ PDFs for three different observational datasets: $\sigma_{\mathrm{LC}} = 0.005$, $\sigma_\mathrm{RV} = 5\,\kms$ (left), $\sigma_{\mathrm{LC}} = 0.015$, $\sigma_\mathrm{RV} = 15\,\kms$ (middle) and $\sigma_{\mathrm{LC}} = 0.025$, $\sigma_\mathrm{RV} = 25\,\kms$ (right). First and last bins hold all other outlying points. Heuristic scan is practically insensitive to the observational data accuracy as long as there are sufficient data points to determine both eclipse depths. Obtained values of temperature ratios are purely statistical and may be compared to the true value of $\tau=0.9452$.} \label{hist_T2T1}
\end{center}
\end{figure}

The weighted values of adjusted parameters are then put into histograms, from which the mean and standard deviation of parameter values are calculated. These estimates are truly statistical, since they do not depend on formal errors of the numerical method. Fig.~\ref{hist_T2T1} shows an example of such histograms for the effective temperature ratio $\tau = T_2/T_1$. Heuristic scan results for this particular example are virtually insensitive to observational data accuracy: for three significantly different cases (labelled \emph{best}, \emph{medium} and \emph{worst} quality data on Fig.~\ref{hist_T2T1}), the outcome of the histogram fit is approximately the same. Histograms for other parameters have somewhat larger standard deviations because of degeneracy: obtained inclination for medium quality data is $85.6^\circ \pm 1.07^\circ$ (compared to the true value $i=85^\circ$) and gravitational potential $\Omega_1$ is $5.44 \pm 0.27$ (compared to the true value $\Omega_1 = 5.244$). It should be noted that reliable statistics implies many starting points of heuristic scan, which in turn implies significant prolongation of the algorithm computation time: each additional scan linearly contributes to the time cost.

Because of data noise and degeneracy, the global minimum is essentially never a single point (with its corresponding uncertainty), it is actually a \emph{region} (with its corresponding uncertainty) in parameter hyperspace. Such a region encompasses many adjacent minima, the depths of which are physically indistinguishable -- a single observed data point with its individual weight may change the identity of the deepest minimum within that region. To identify these regions, \phoebe computes \emph{convergence tracers} -- selected 2D cross-sections of the parameter hyperspace, tracing parameter values from each starting point, iteration after iteration, all the way to the converged solution. Attractors -- regions that attract most convergence traces -- within these cross-sections reveal parameter correlations and degeneracy. Inspecting such convergence tracers offers additional insight on the quality and integrity of the solution. Local minima in context of convergence tracers are those that lay outside of the deepest attractor(s); those are the ones that need to be identified and escaped from.

A particularly troublesome degeneracy is the one between the inclination and either of the effective potentials $\Omega_{1,2}$ of the two stellar components (which act on behalf of components' radii). Fig.~\ref{tracer}(a) shows the $i$-$\Omega_1$ convergence tracer computed for our test binary. The correlation between $i$ and $\Omega_1$ is evidently very flat at $i \sim 85^\circ$, which may be easily understood: the model is able to compensate smaller inclinations by enlarging the radius of the star and vice versa. Therefore we should not trust light curve analysis to disentangle these parameters by itself -- additional constraints are needed. This issue will be further discussed in Section \ref{physics}.

\subsection{Parameter Kicking} \label{parameter_kicking}

Another possible approach to detect and escape from local minima is to use a stochastic method such as Simulated Annealing (SA). However, such methods are notoriously slow. Thus, instead of full-featured SA scan, a simple new procedure has been developed that achieves the same effect as stochastic methods, but in significantly shorter time. The idea is as follows: whenever a minimum is reached within a given fractional accuracy, the algorithm runs a globality assessment on that minimum. If we presume that standard deviations $\sigma_k$ of observations are estimated properly and that they apply to all data points, we may use them for $\chi^2$ weighting:
\begin{equation}
\chi_k^2 = \sum_{i=1}^M w_k w_i (x_i - y_i)^2 = \frac 1{\sigma_k^2} \sum_{i=1}^M w_i (x_i - y_i)^2,
\end{equation}
where index $i$ runs over $M$ measurements within a single data-set and index $k$ runs over $N$ data-sets (photometric and RV curves); $x_i$ are observed data points, $y_i$ are calculated data points and $w_i$'s are individual weights. Since the weighted variance is given by:
\begin{equation}
s_k^2 = \frac{1}{N_k-1} \sum_i w_i (x_i - y_i)^2,
\end{equation}
we may readily express $\chi_k^2$ as:
\begin{equation}
\chi_k^2 = (N_k - 1) \frac{s_k^2}{\sigma_k^2}.
\end{equation}
and the overall $\chi^2$ value as:
\begin{equation}
\chi^2 = \sum_k (N_k - 1) \left( \frac {s_k}{\sigma_k} \right)^2.
\end{equation}

If $\sigma_k$ are realistic, the ratio $s_k / \sigma_k$ is of the order unity and $\chi^2$ of the order $N_\mathrm{tot} = \sum_k N_k$. This we use for parameterizing $\chi^2$ values:
\begin{equation} \label{eq_lambda}
\lambda := \left( \chi^2 / N_\mathrm{tot} \right) \,: \textrm{quantization.}
\end{equation}

\begin{figure}
\begin{center}
\includegraphics[width=12cm, height=6cm]{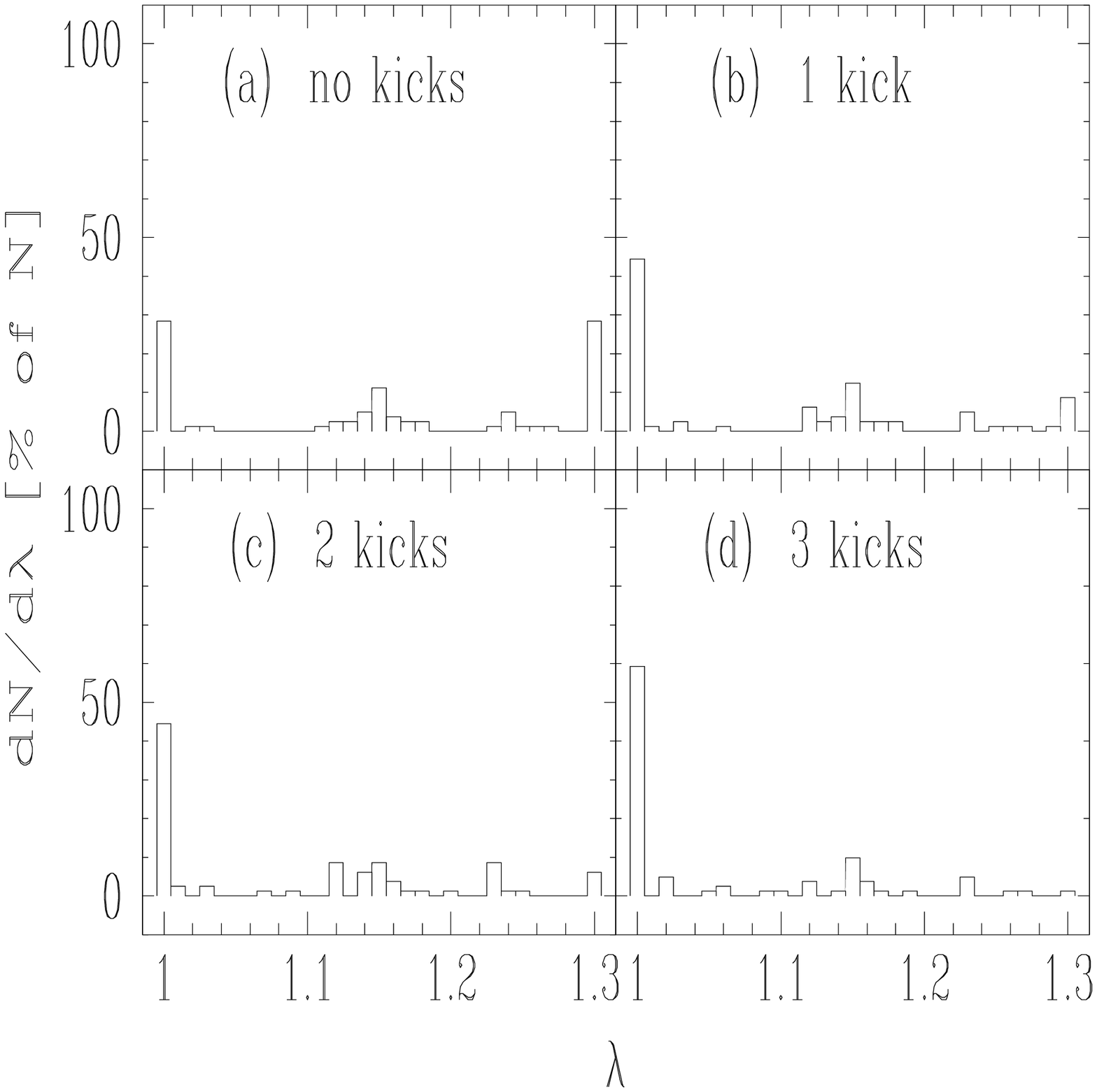}
\caption{$\lambda$-histogram for initial heuristic scan and three consecutive parameter kicks. The success of parameter kicking is obvious, since after only three consecutive kicks the percentage of scans that converge within 1\% of the optimal value of $\lambda$ (in case of proper $\sigma_k$'s $\lambda = 1$) is \emph{doubled} from $\sim$30\% to $\sim$60\%. As is shown by \cite{prsa2005}, parameter kicking proves to be even more efficient in case of exclusively photometric observations, with improvement from $\sim$15\% to $\sim$75\% in convergence after three consecutive kicks.} \label{lambda}
\end{center}
\end{figure}

\begin{figure}
\begin{center}
\includegraphics[width=16cm]{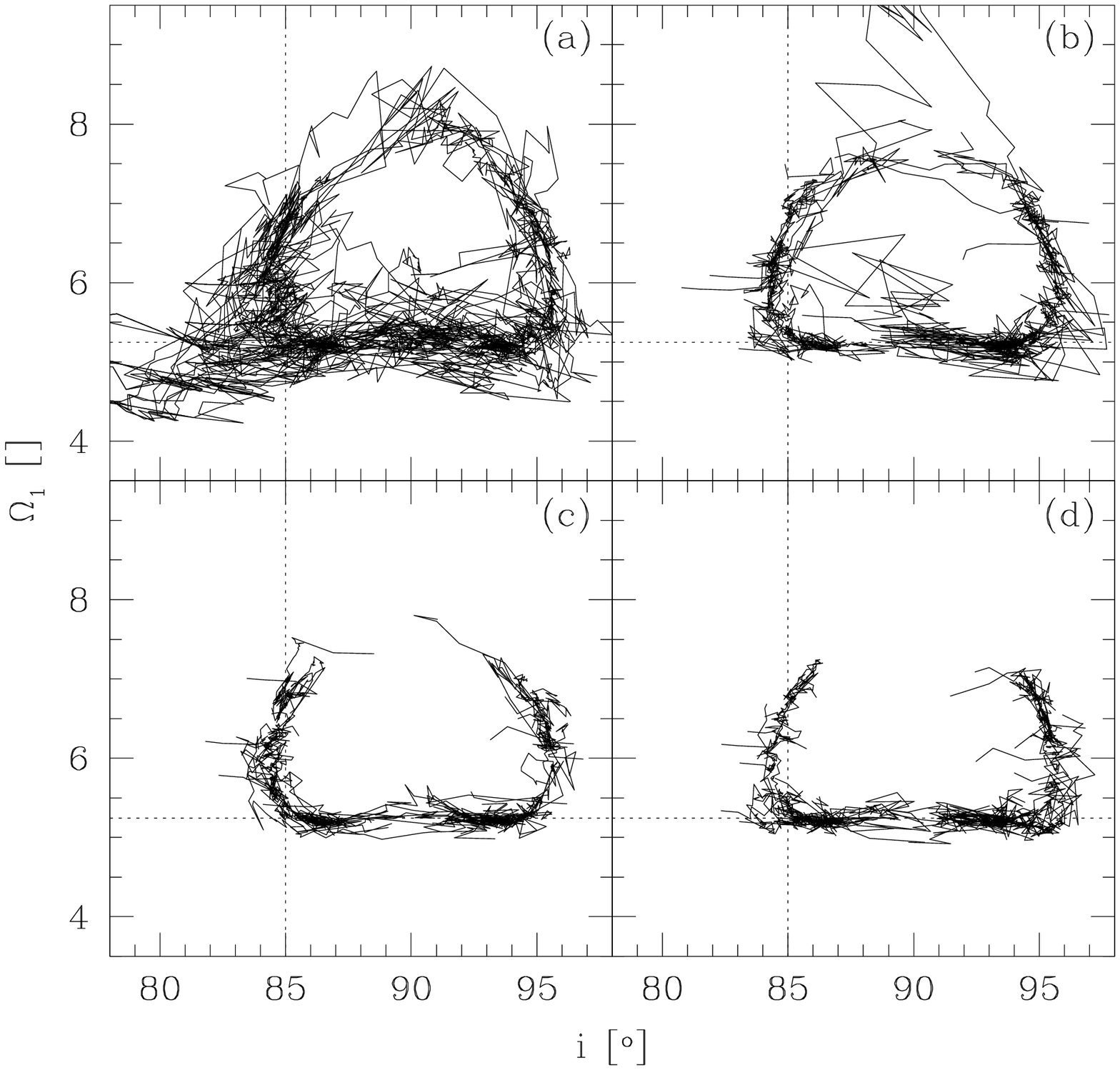}
\caption{Convergence tracer for $i$-$\Omega_1$ cross-section. This particular case is a notorious example of very difficult-to-handle correlation between the inclination and effective potentials (hence the radii) of both components (only $\Omega_1$ correlation is depicted for brevity). Individual plots denoted with letters (a) through (d) show the result of NMS heuristic scan from zero to three consecutive parameter kicks. Cross-hairs mark the position of the true minimum. Attractors are symmetric to $i=90^\circ$, but still very flat at $i \sim 85^\circ$ to $90^\circ$ interval, which means that the obtained NMS solution should not be blindly trusted; rather, additional constraining is needed.} \label{tracer}
\end{center}
\end{figure}

Parameter kicking is a way of knocking the obtained parameter-set out of the minimum: using the Gaussian PDF, the method randomly picks an offset for each parameter. The strength of the kick is determined by the Gaussian dispersion $\sigma_{\mathrm{kick}}$, which depends on the minimum globality assessment parameter $\lambda$. If $\lambda$ is high, then the kick should be strong, but if it is low, i.e.~around $\lambda \sim 1$, then only subtle perturbations should be allowed. Experience shows that a simple expression such as:
\begin{equation}
\sigma_{\mathrm{kick}} = \frac{0.5 \lambda}{100}
\end{equation}
works very efficiently in case of partial eclipses. This causes $\sigma_{\mathrm{kick}}$ to assume a value of $0.5$ for $10 \sigma$ offsets and $0.005$ for $1 \sigma$ offsets, being linear in between. Note that this $\sigma_{\mathrm{kick}}$ is \emph{relative}, i.e.~given by:
\begin{equation} \label{rel_to_abs}
\sigma_{\mathrm{kick}}^{\mathrm{abs}} = x \, \sigma_{\mathrm{kick}}^{\mathrm{rel}},
\end{equation}
where $x$ is the value of the given parameter. When convergence within the given fractional accuracy is reached, parameters are kicked with respect to the depth of the minimum and the minimization is restarted from displaced points. The influence of consecutive parameter kicking with NMS is depicted in Fig.~\ref{lambda}; it is shown that out of all heuristic scans only $\sim$30\% initially converge to within 1\% of optimal value of $\lambda$, whereas this percentage steadily grows to $\sim$60\% after three kicks. Figs.~\ref{tracer}(b)--(d) show significant improvement to the solution introduced by these consecutive kicks. Parameter kicking is able to quickly escape from local minima and thus rapidly increase convergence efficiency of the whole NMS method. A down-side of parameter kicking is the time cost: each additional kick linearly adds to the overall execution time. For thorough discussion and details on benchmarking please refer to \phoebe accompanying documentation.

The idea behind the NMS implementation is \emph{not} to replace DC, but rather to \emph{complement} it. DC is created for interactive usage and converges in discrete steps that need monitoring. NMS on the other hand aims to automate this process so that intermediate monitoring is no longer necessary. DC is one of the fastest methods ({\sf WD}'s {\tt DC} in particular, since it is optimized for EBs), but may easily diverge. At the expense of speed, NMS is one of the most robust algorithms for solving non-linear minimization problems and never diverges. Finally, both DC and NMS suffer from degeneracy and may become stuck in local minima. To overcome this, both methods are complemented by heuristic scan and parameter kicking. These differences in intent make a combination of the two methods a powerful engine for solving the inverse problem.

\section{Extended set of physical constraints} \label{physics}

{\sf WD}'s extensive list of more than 30 adjustable parameters is an overwhelming indicator of how sophisticated the model has become in 35 years of development. Nevertheless, accuracy is crucial for a model to describe such a wide diversity of intrinsically different binaries. An accurate model should contain all relevant physical contributions for which the governing laws are well-known. We start the discussion by introducing new physical ties and constraints to parameter extracting schemes that are implemented in \phoebents. It should be stressed that all these constraints are \emph{optional} and it is up to the user to select the ones that are of interest.

\subsection{Color indices as indicators of individual temperatures}

One of the main difficulties of modeling EBs is accurate determination of individual temperatures of both components. Frequent practice in literature is to \emph{assume} the temperature of one star (e.g.~from spectra or color indices) and fit the temperature of the other star. This approach is often inadequate, particularly for binaries with similar component temperatures and luminosities: in such cases, the contribution of both components to the system luminosity is significant and it is difficult to accurately estimate the contribution of only one star in advance.

Before we propose a method capable of providing individual temperatures from standard photometry observations without any \`a priori assumptions, it proves useful to introduce a concept of effective temperature \emph{of the binary}. A binary may be regarded as a point-source, the effective temperature of which varies in time. To this effective temperature contribute both components according to their sizes and individual temperatures, and the inclination. Effective temperature of the binary is directly revealed by the color index, so its observational behavior is well known. If a model is to accurately reproduce observations, the composite of contributions of both components must match this behavior.

The observational light curve quantity (dependent variable) {\sf WD} works with is \emph{flux}, scaled to an arbitrary level (which could also be in absolute physical units, i.e.~W/m${}^2$ per wavelength interval). The model adapts to this level by determining the corresponding passband luminosity $\hla$, one for each passband. However, these passband luminosities are completely decoupled from one another, so any color information that might have been present in the data is discarded. Since the effective temperature of a binary is observationally revealed by its $B-V$ (or any other suitable) color index\footnote{Useful relations among color indices are given in \cite{caldwell1993}.}, some of the relevant temperature information is lost. Transformation to fluxes in absolute units would not suffice for properly determining the corresponding passband luminosities -- one needs a \emph{physical relation} between those luminosities. Neglecting this additional relation may result in discordant colors between the temperatures obtained by the fit (assuming that $T_1$ is \`a priori known) and the ones determined by the binary's effective temperature. This relation is nothing else than the color index and may thus be accurately determined from observations.

In the last decade substantial effort was made to scan the sky for standard stars to be used for photometric calibration: \cite{landolt1992} covering celestial equator, \cite{henden1997} and \cite{bryja1999} covering fields around cataclysmic variables, \cite{henden2000} covering fields around symbiotic binaries, to name just a few. These efforts help overcome the problem of small CCD fields with respect to all-sky photometry, since in many fields there are now cataloged standard stars that may be used to extract color indices for EBs. In context of \phoebents, this means that using measured color indices as additional information is plausible even if the data were not obtained under photometric conditions.

\phoebe initially regards $\hla$'s as simple level-setting quantities -- physical context comes in only after the color index constraint is set. For the sake of simplicity, consider that input observational data are supplied in magnitudes (such that the color indices are meaningful) rather than fluxes. \phoebe input data in individual passbands should not be scaled arbitrarily; that is \phoebents's job.

The native type that \phoebe works with is inherited from {\sf WD}, which is flux. \phoebe uses a single, passband-independent parameter $m_0$ to transform \emph{all} light curves from magnitudes to fluxes. The value of $m_0$ is chosen so that the fluxes of the dimmest light curve are of the order of unity. It is \emph{a single quantity} for all light curves, which immediately implies that the magnitude difference, now the flux ratio, is preserved; hence, the color index is preserved. If the distance to the binary is known (e.g.~from astrometry), $m_0$ immediately yields \emph{observed} luminosities of the binary; intrinsic luminosities are obtained if the color excess $E(B-V)$ is also known.

This is where physics comes in: from such set of observations, the calculated $\hla$'s are indeed passband luminosities, the ratios of which are the constraints we need: passband luminosities of light curves are now connected by the corresponding color indices. Once the color constraint is set, \phoebe makes sure that the ratio between $\hla$'s is kept constant. 

Now that the color indices are preserved, effective temperature of the binary may be obtained from a color--temperature calibration. \phoebe uses updated \cite{flower1996} tables with coefficients given in Table \ref{flower_cmd}. It should be stressed that the color constraint is applicable only if the data are acquired on (or properly transformed to) a standard photometric system.

\begin{deluxetable}{crr}
\tablewidth{0pt}
\tablecaption{\label{flower_cmd} Coefficients of the empirical $\Teff$~$(B-V)$ relation given by the 7th degree polynomial fit $\Teff = \sum_{i=0}^7 C_i (B-V)^i$ (Flower, private communication). The second column applies to main-sequence stars, sub-giants and giants, the third column applies to supergiants.}
\tablehead
  {
  \colhead{Coefficient:} & \colhead{\sf V, IV, III, II} & \colhead{\sf I}
  }
\startdata
$C_0$ &  3.979145 &  4.012560 \\
$C_1$ & -0.654992 & -1.055043 \\
$C_2$ &  1.740690 &  2.133395 \\
$C_3$ & -4.608815 & -2.459770 \\
$C_4$ &  6.792600 &  1.349424 \\
$C_5$ & -5.396910 & -0.283943 \\
$C_6$ &  2.192970 & $-$       \\
$C_7$ & -0.359496 & $-$       \\
\enddata
\end{deluxetable}

Applying the color constraint, effective temperatures of individual components may be readily disentangled by the minimization method. The method is now able to find only those combinations of parameters that \emph{preserve} effective temperature of the binary and hence the color index. Since the relation between effective temperatures of individual components is fully determined by the light curve shape (dominantly by the primary-to-secondary eclipse depth ratio) and since the sum of both components' contributions must match the effective temperature of the binary, the color-constrained minimization method yields effective temperatures of individual components without any \`a priori presumptions.

Let us demonstrate this concept on our test binary. Calculating passband luminosities from medium quality observations ($\sigma_{\mathrm{LC}} = 0.015$, $\sigma_\mathrm{RV} = 15\,\kms$) yields $L_B/L_V = 0.592 \pm 0.006$. Transforming this into magnitudes yields the color index $B-V = 0.57 \pm 0.01$, which in turn yields effective temperature of the binary to be $T_\mathrm{eff} = 6\,002\,\,\mathrm K \pm 40$\,K. The relation between both individual temperatures from the ratio of eclipse depths is well determined (c.f.~Fig.~\ref{hist_T2T1} yielding $T_2 / T_1 = 0.946 \pm 0.003$), disentangling effective temperatures of individual stars to be $T_1 = 6\,190\,\,\mathrm K \pm 52$\,\,K and $T_2 = 5\,880\,\,\mathrm K \pm 51$\,\,K. Comparing these values to true values $T_1 = 6\,200$ K and $T_2 = 5\,860$ K is very encouraging.

\subsection{Spectral energy distribution as independent data source}

Traditionally, spectral energy distributions (SED) have been used only indirectly, e.g. for extracting radial velocities or determining effective temperatures. Several recent studies of individual EBs have shown that including flattened SEDs may be used as external check of the model solution (see \cite{siviero2004} and \cite{marrese2004} for examples), where individual spectral lines of \'Echelle spectra are compared with \citet{kurucz1998} model atmospheres.

Since the Kurucz's model atmosphere program runs only under VAX/VMS in its distributed form, several databases of precomputed spectra have been assembled for practical use (e.g.~\cite{zwitter2004} covering the spectral range 765--875nm, \cite{murphy2004} covering 300--1000nm, \citet{munari2004b} covering 250--1050nm). Such databases bring stellar atmospheres to non-VAX/VMS equipped users. In recent years significant effort has been made to port Kurucz's model atmospheres code to Linux (see e.g.~CCP7 initiative at {\tt http://www.stsci.edu/software/CCP7}, \cite{sbordone2004} and others). Such initiatives enable users to include SED data in solving the inverse EB problem. \phoebe already takes a step in that direction by using a synthetic spectra database to test whether flattened, wavelength-calibrated spectra match synthetic spectra within a given level of significance.

One very important caveat that should be stressed: it is not feasible to compare observational SEDs to synthetic SEDs over the full spectral range. The problems occur because of Earth's atmosphere (significant parts of the spectrum are dominated by telluric lines, which the model does not handle). By default, \phoebe uses the \cite{zwitter2004} grid of 61\,196 synthetic spectra covering the 765--875 nm interval at a resolving power $R=20\,000$. A simple interpolation may be used to obtain the spectrum characterized by any combination of $\Teff$, $\logg$, [M/H] and $\vrot$ with the accuracy better than $25$ K in temperature, $0.05$ dex in $\logg$ and metallicity and 1 $\kms$ in rotational velocity. These uncertainties are smaller than the uncertanties of the Kurucz's model for parameters of our test binary, so interpolation does not induce any systematic errors.

\begin{figure}
\begin{center}
\includegraphics[width=16cm, height=8cm]{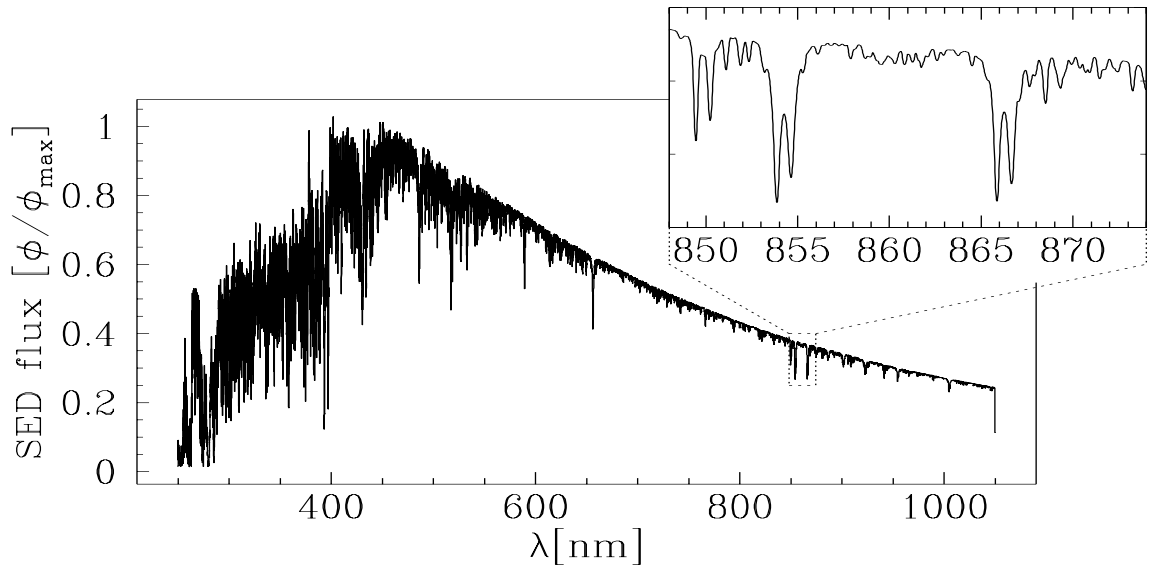}
\caption{Synthetic spectrum of the test binary at quarter phase. The spectrum is built by Doppler-shifting and convolving individual component spectra obtained by linear interpolation in $T_\mathrm{eff}$ and $\logg$ from precomputed stellar spectra tables by \citet{munari2004b}. The inset magnifies a part of the spectrum corresponding to the Gaia RVS wavelength range, which is covered by the \cite{zwitter2004} database. The strongest lines in the inset are split (revealing the binary nature of the object) and are due to Ca II (849.80\,\,nm, 854.21\,\,nm and 866.94\,\,nm).} \label{sed}
\end{center}
\end{figure}

To demonstrate current level of SED implementation in \phoebents, consider again our test binary. Parametric vectors ($\Teff$, $\logg$, $\vrot)_{1,2}$ of both EB components are determined by the model solution from photometric and RV data. These are used to obtain synthetic spectra by linear interpolation in $T_\mathrm{eff}$, $\logg$ and $v_{\mathrm{rot}}$ from the grid. For the "true" simulated spectrum, solar abundances ([M/H] $=0.0$), corotation ($v_{\mathrm{rot}_1} = 64 \kms$, $v_{\mathrm{rot}_2} = 52 \kms$) and microturbulence $v_\mathrm{turb} = 2 \kms$ are assumed. Effective spectrum of the binary is computed in out-of-eclipse phase (e.g. quarter-phase) by Doppler-shifting and convolving the spectra of the two stars. Fig.~\ref{sed} shows an example of a quarter-phase spectrum of the test binary.

\begin{figure}
\begin{center}
\includegraphics[width=16cm]{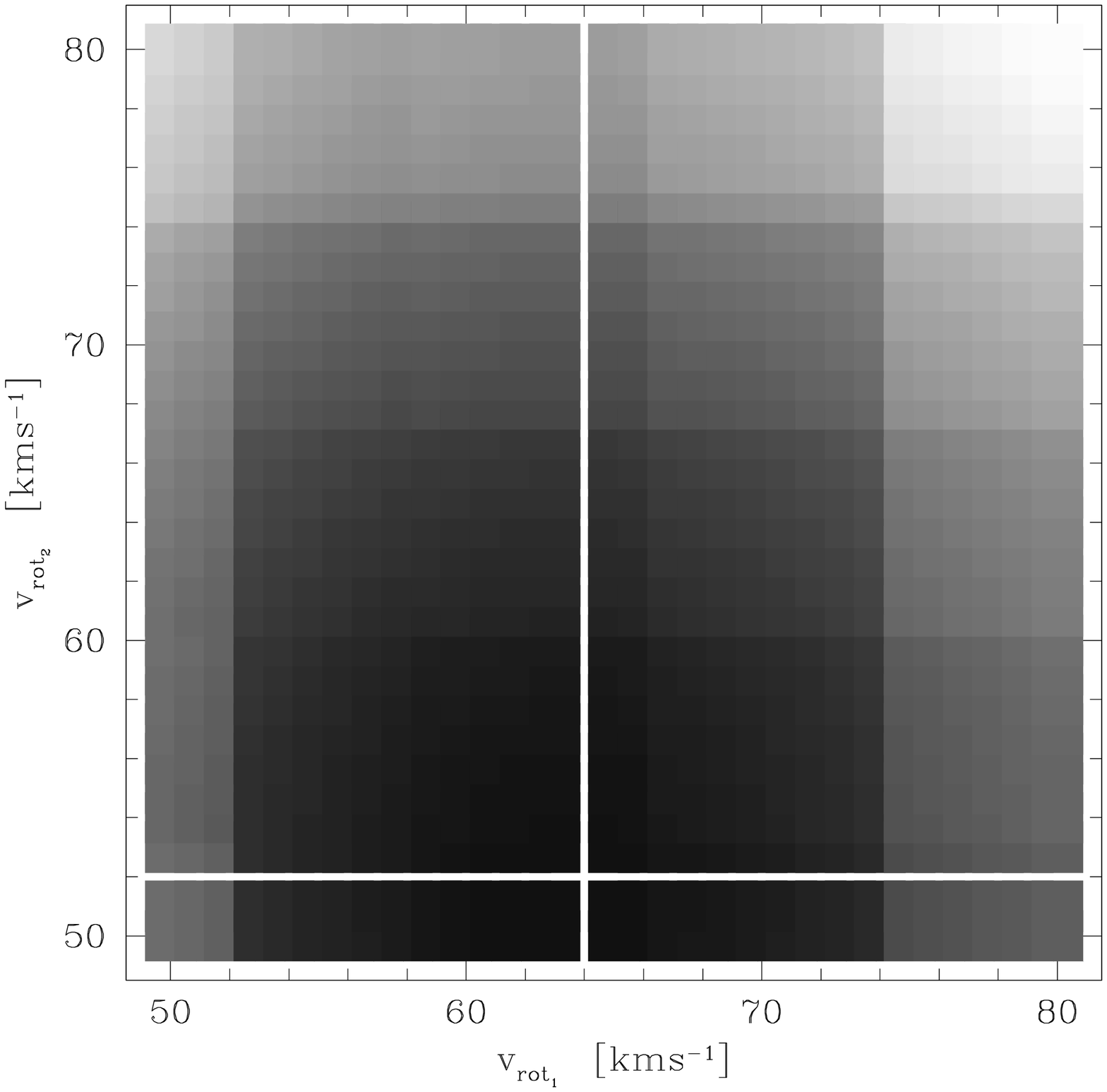}
\caption{Result of the $\chi^2$ comparison of the "true" spectrum against \cite{zwitter2004} database. Out of all cross-sections, $v_{\mathrm{rot}_1}$-$v_{\mathrm{rot}_2}$ cross-section is most interesting, because it helps break the degeneracy between effective gravitational potentials $\Omega_1$ and $\Omega_2$. The levels of gray in the mesh are linear in $\chi^2$ and denote the quality of the fit: black color corresponds to the best fit, white color corresponds to the worst fit. Cross-hairs denote the position of the true values of rotational velocities.} \label{sed_mesh}
\end{center}
\end{figure}

After each solution of the NMS heuristic scan, a synthetic spectrum is built\footnote{At present the spectrum may be generated for any orbital phase outside of eclipses.} from that solution. It is then compared with the "true" spectrum by the $\chi^2$ cost function. Effective temperature is the dominant parameter that governs SED shape, but this is of little use for our case: recall from Fig.~\ref{hist_T2T1} and the discussion on color indices that individual temperatures are well determined from photometry alone. Rather, our solution suffers from degeneracy in effective potentials $\Omega_1$, $\Omega_2$ and inclination $i$ (Fig.~\ref{tracer}). It would be beneficial if the SEDs could break this degeneracy. Since the mass ratio and semi-major axis of the model are effectively held constant by the RVs, $\Omega_1$ and $\Omega_2$ depend only on the radii of individual components. Thus, different $\Omega$'s imply different $\logg$ and, by assuming corrotation, also $\vrot$. Fig.~\ref{sed_mesh} shows the $\vrot{}_1$-$\vrot{}_2$ cross-section, demonstrating that, as we hoped, the SED analysis indeed \emph{constrains} the solution to smaller intervals for $\vrot{}_1$ and $\vrot{}_2$, thus smaller intervals for $\Omega_1$ and $\Omega_2$.

The $\vrot{}_1$-$\vrot{}_2$ cross-section may sometimes do even more than only break the degeneracy between $\Omega_1$ and $\Omega_2$. If the radii are well determined, e.g.~by total eclipse geometry, such analysis yields synchronicity parameters $F_1$ and $F_2$, since the only way to compensate the change in rotational velocities for any predetermined radii is to break the corotation presumption. This may be especially important in analysis of well detached systems, as demonstrated by \cite{siviero2004}.

It should be noted that there is \emph{no support} for extracting $\Teff$, $\logg$, [M/H] or $\vrot$ from spectra at the moment, only a weighted $\chi^2$ test is done to confirm or reject the particular set. As such, the current implementation forms the base of spectral analysis for EBs, but it still does not contribute fully to minimization. Once we are capable of building stellar spectra without presuming spherically-symmetrical stars in LTE, full SED will be introduced to the minimization process as well. Such a scheme will have to weight properly individual wavelengths, since there is much less information in the continuum of the spectrum than it is, in example, in central parts and wings of spectral lines. However, even the present implementation of SED analysis finds the values of physical parameters which have not usually been attainable by light and RV curve analysis, namely metallicity and rotational velocity \citep[see][]{terrell2003}.

\subsection{Main sequence constraints}

In cases where SED observations are not available, or where they are used only to extract RVs, the degeneracy among parameters may still be so strong that neither heuristic scan nor parameter kicking can break it. In such cases we stand no chance of obtaining \emph{any} satisfactory solution without further constraining the modeled binary.

{\sf WD} features 8 modes of operation that determine the morphology of the binary. By deciding on the mode of operation, the user imposes a set of physical constraints; for example, both components of over-contact systems have equal potentials. \phoebe refers to these constraints as \emph{morphological} constraints. If a morphological constraint is not chosen properly, the model may converge to a physically implausible solution.

On the other hand, we can sometimes make an assumption, not being certain it is correct. In case of degeneracy, a solution based on an assumption may be better than having no solution at all. One assumption might be the age of the coeval components. Assuming a particular type of evolutionary track, the luminosities from stellar evolution models may then be obtained \citep{pols1995}. Another such assumption could be the distance to the binary, e.g.~from astrometry. Yet another assumption may be that either or both components are main-sequence stars. Since a significant percentage of all stars are on the main-sequence, there is a fair chance that our assumption is correct.

Applying main-sequence constraint to component(s) of the modelled binary means imposing $M$-$L$-$T$-$R$ relations for main-sequence stars (see e.g.~\cite{malkov2003} for such relations specific to EBs). Consequentially, given a single parameter (e.g. component's effective temperature), all other parameters (its mass, luminosity and radius) are calculable. This in turn implies that, in case of circular and nearly-circular orbits, effective potential of the constrained component is fully determined. Main-sequence constraint may be used for testing whether either or both stars may plausibly be main-sequence stars: depending on behavior of the $\chi^2$ value, such hypothesis may be accepted or rejected.

Such additional constraints are not as straight-forward as was the case with morphological constraints. For example, by implying the condition: \emph{let the modeled binary be a main-sequence binary}, we break the degeneracy by selecting the one solution that corresponds to that condition. This is why \phoebe refers to these constraints as \emph{conditional} constraints (CC). It is very important to emphasize that using conditional constraints improperly may lead to creating and propagating a circular argument: EBs provide absolute parameters for stars, which can then be used to establish various calibrations. Conditional constraints on the other hand use calibrations to constrain derived parameters. Conditionally constrained solutions should thus never be used to establish calibrations of any kind.

\begin{figure}
\begin{center}
\includegraphics[width=8cm]{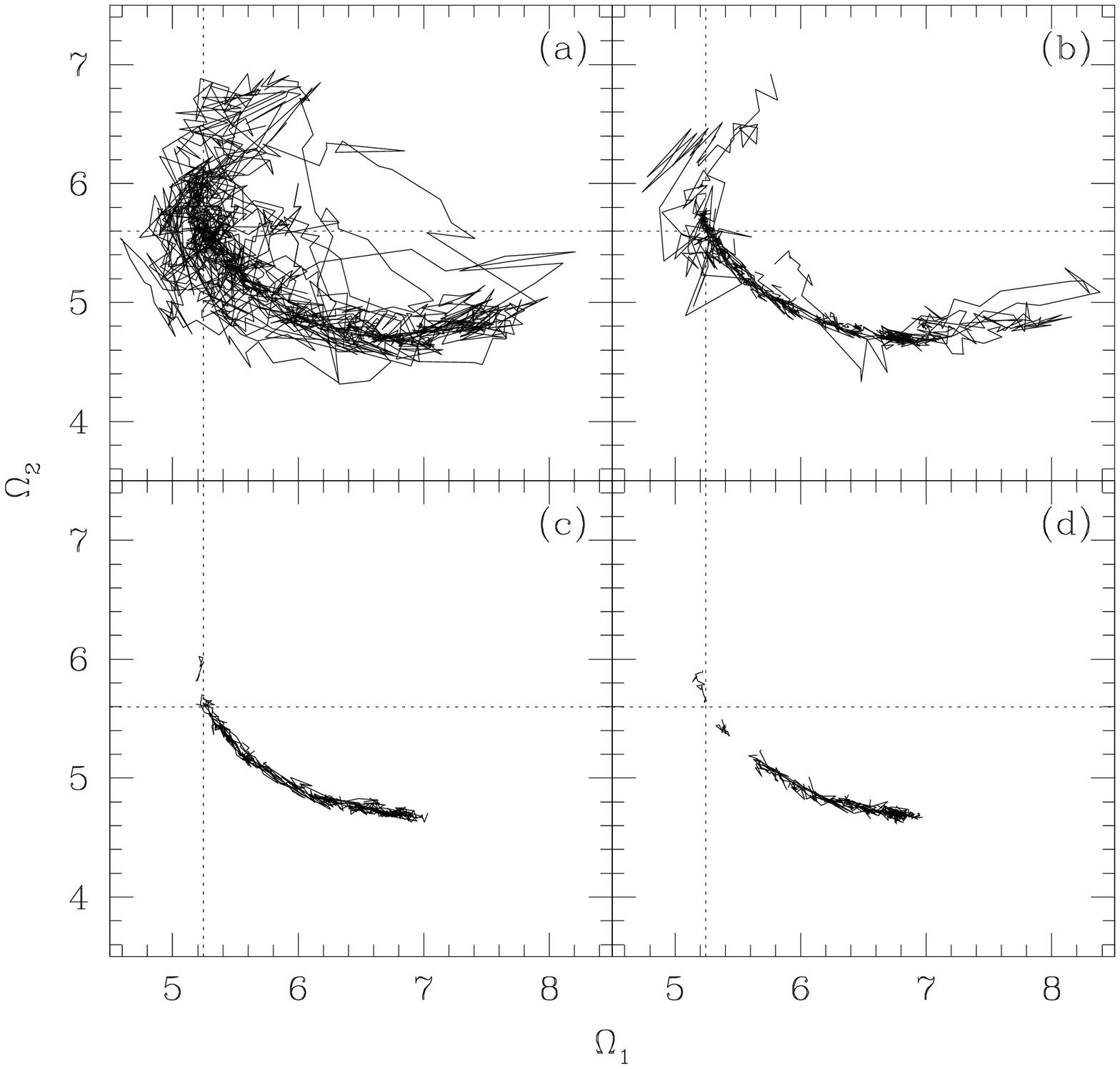}
\includegraphics[width=8cm]{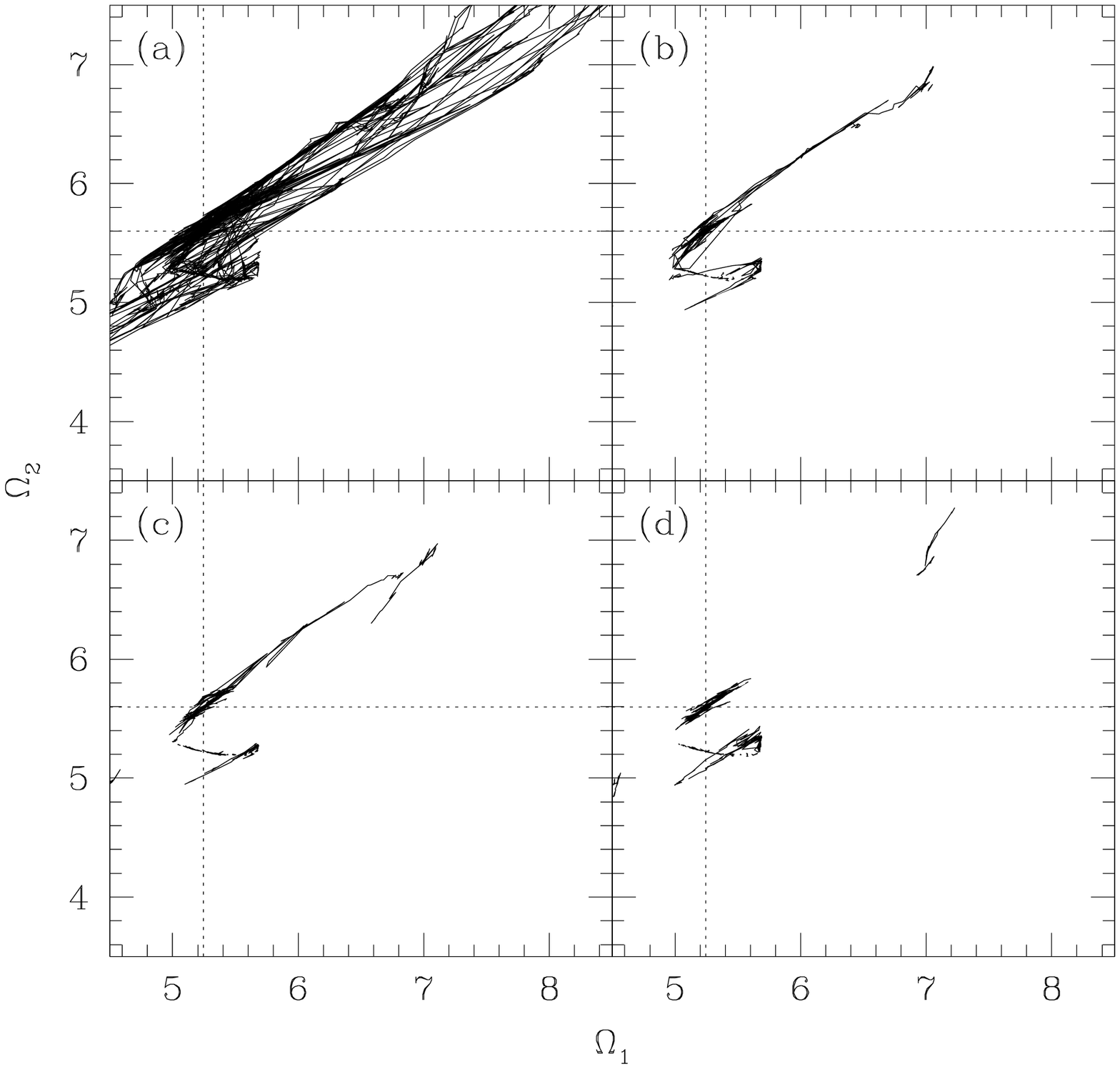} \\
\caption{Convergence tracers for $\Omega_1$-$\Omega_2$ cross-section without (left) and with (right) the main-sequence constraint imposed on the model. Similar to Fig.~\ref{tracer}, panels (a) through (d) denote successive number of kicks (zero to three) and cross-hairs mark the correct solution. Comparing these results immediately shows that the intersection of the two attractors yields the correct solution. Note that there are two intersections because of the model symmetry to the labeling of the two components (primary and secondary roles of both stars may be interchanged).} \label{msc}
\end{center}
\end{figure}

Recall from Fig.~\ref{tracer} that the solution from photometric and RV observations of our test binary indeed suffered from degeneracy in inclination and potentials. If we conditionally constrain both modelled components with the main sequence constraint, potentials $\Omega_1$ and $\Omega_2$ are calculable and thus exactly known. The variation in their values is only a consequence of the variation in either of the main-sequence parameters ($M$, $L$, $T$ or $R$) that accomodate for different orbital inclinations during the fit. Fig.~\ref{msc} shows convergence tracers for a similar NMS heuristic scan as in Fig.~\ref{tracer}, this time for $\Omega_1$-$\Omega_2$ cross-section without (left) and with (right) the main-sequence constraint imposed on the model. Since the main-sequence constraint is very strong, there is no practical need for heuristic scan or parameter kicking (both $\Omega_i$'s are calculable for the given inclination and convergence is thus assured from practically \emph{any} point in the hyperspace); Fig.~\ref{msc} (right) still depicts both heuristic scan and consecutive parameter kicks for comparison between convergence tracer shapes and slopes of unconstrained and main-sequence constrained model. It is evident that both solutions intersect, yielding the right solution. This is of course expected, since our test binary is in fact composed of two main-sequence components.

One would hope that total eclipses reduce the degeneracy, but this also does not necessarily happen. If stars have comparable sizes (as is the case of our test binary), the duration of the eclipse totality is very short and limb darkening may obscure its flatness. Since geometrical parameters in case of total eclipses are better constrained (the corresponding hyperspace cross-sections feature very narrow valleys), parameter kicking may work to our disadvantage, knocking the solution far from the minimum by only a small parameter displacement. This issue will be addressed in detail in the follow-up paper.

By using conditional constraining, we select a preferred subspace of model solutions. This is why extra care should be taken for the choice of adopted CCs.

\subsection{Interstellar and atmospheric extinction} \label{interstellar_extinction}

Although interstellar extinction has been discussed in many papers and quantitatively determined by dedicated missions (IUE, 2MASS, and others), the approach for EBs is often inadequate. Reddening is usually calculated by analytic approximation \citep[e.g.][]{lang1992} or extinction tables \citep[e.g.][]{schlegel1998}, using EB's galactic coordinates and inferred distance to the binary. Such calculations are performed only for a single, effective wavelength of the given passband; the obtained value is then subtracted uniformly from all photometric observations in that passband.

If the amount of reddening is not negligible and the binary components have significantly different surface temperatures, effective temperature of the binary is a function of phase because of eclipses. This is why in case of strongly reddened EBs it cannot be assumed that the correction due to reddening is well approximated by simply subtracting a constant. A dedicated study of this problem has been presented by \cite{prsa2004}, here we only overview the conclusions.

\phoebe uses already described synthetic SED data for rigorous reddening corrections. It builds an intrinsic effective spectrum of the binary by Doppler-shifting and convolving the spectra of individual components as a function of phase. This intrinsic spectrum is then rigorously (wavelength-by-wavelength) reddened by the formula proposed by \cite{cardelli1989} and convolved with the given passband transmission function. Photometric magnitude is then obtained by integrating the reddened spectrum over the given passband.

\begin{figure}
\includegraphics[width=8cm,height=4cm]{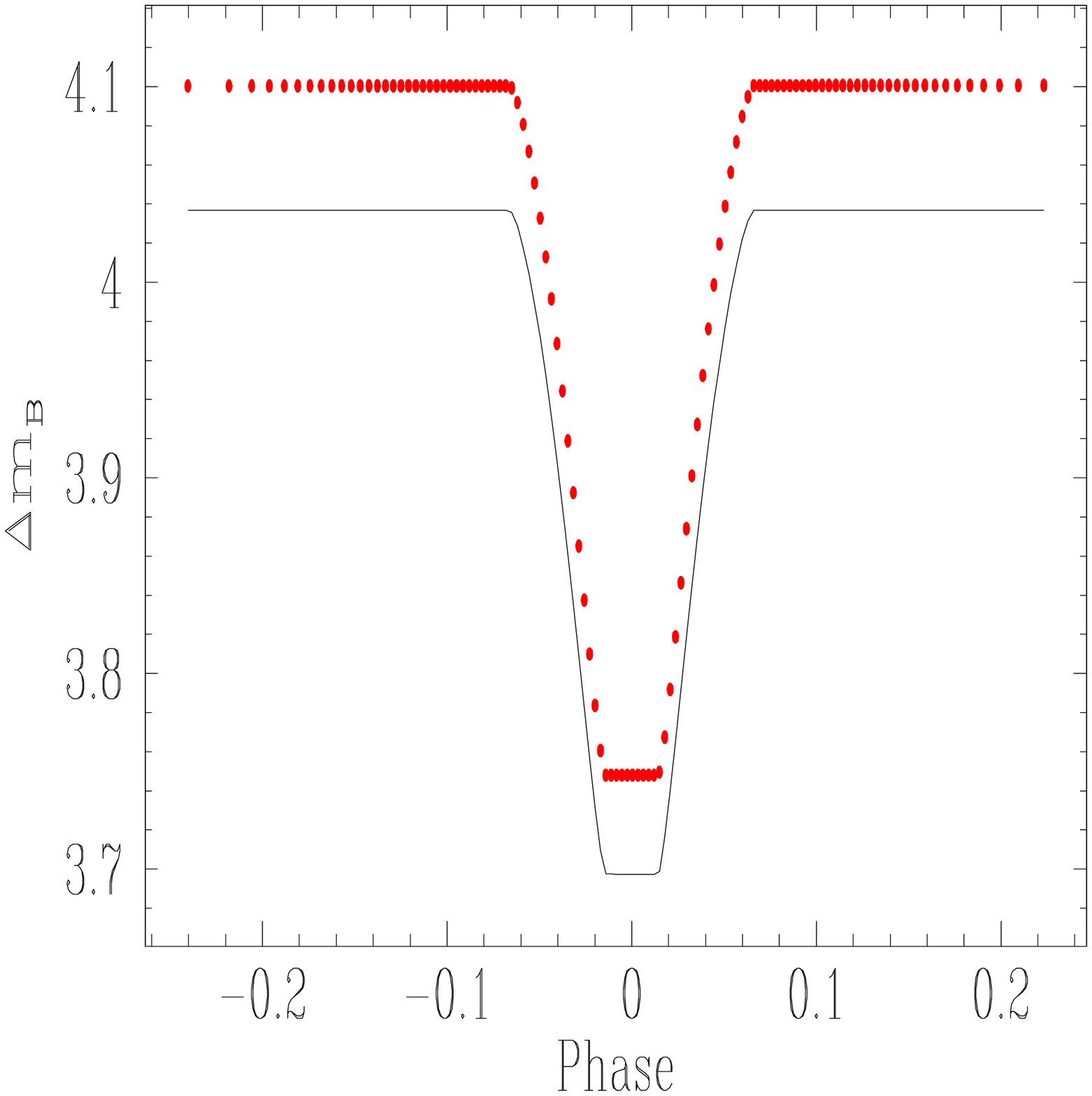}
\includegraphics[width=8cm,height=4cm]{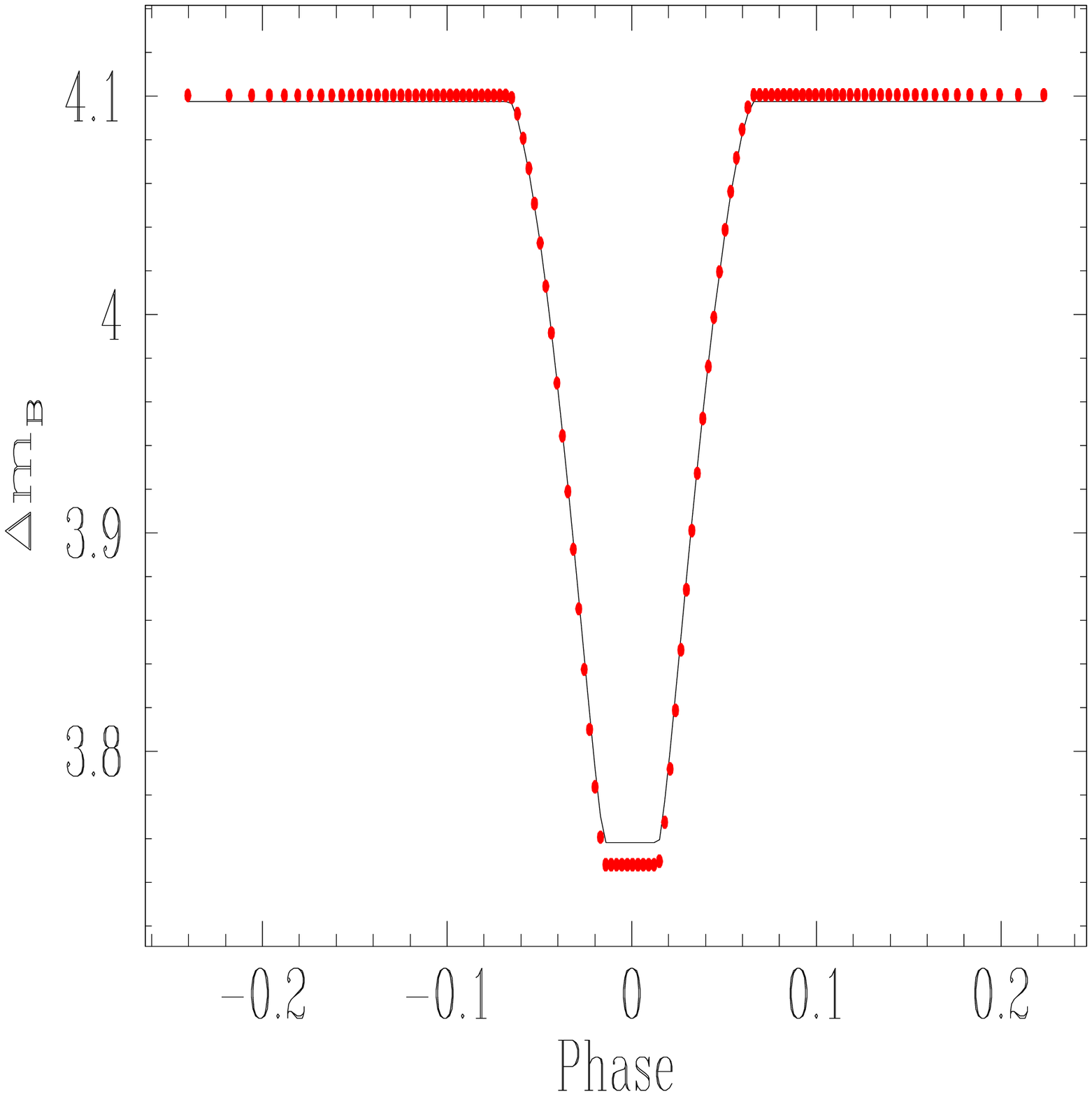} \\
\caption{Johnson B magnitude difference between reddened and unreddened observations. $E(\mathrm B - \mathrm V) = 1$ and $i=90^\circ$ is assumed with other parameters listed in Table \ref{synthetic_binary}. Left: The discrepancy between the rigorously applied reddening (points) and the constant subtraction approach (solid line). The difference due to erroneous approach is $\sim 0.06$ mag. Subtracted constant was obtained from the effective wavelength ($\lambda_{\mathrm{eff}} = 4410.8$~\AA) of the Johnson B passband transmission curve. Right: Overplotted light curves with the subtraction constant calculated properly in out-of-eclipse regions. There is still a \emph{measurable} difference of $\sim 0.01$~mag in eclipse depth of both light curves. Adapted from \citet{prsa2004}.} \label{reddening}
\end{figure}

There are two major implications of this improved scheme over the traditional constant subtraction, that are depicted in Fig.~\ref{reddening}: {\bf 1)} using effective passband wavelength to calculate the reddening constant introduces a significant systematic error in reproduced magnitudes. Effective wavelength of the passband is irrelevant for the reddening: it is the spectrum integral over the passband interval which must be the same in both approaches. {\bf 2)} Even if the constant was calculated properly (by making sure that the integrals over the passband are the same), there would still be a measurable offset in both eclipses due to the change in effective temperature of the binary. This effect gains on significance as the temperature difference between both components grows and may reach $\sim 0.2$ mag or more in case of symbiotic binaries \citep{prsa2004}.

Atmospheric extinction has a similar effect on photometric observations, since it also reddens the data as a function of wavelength. The study has shown that interstellar extinction dominates the whole wavelength range in cases of strong reddening, while atmospheric extinction dominates the blue parts of the spectrum in cases of low-to-intermediate reddening \citep{prsa2004}. The main difference between interstellar and atmospheric extinction is that the latter is usually taken into account during initial reduction of the photometric data, so prior to running \phoebents.

\section{Conclusion} \label{conclusion}

This paper has presented current stage of an ongoing effort to unify proven ideas and new approaches of the EB field. \phoebe is built on top of the {\sf WD} code and decades of experience put into its development, trying to add its own pieces to the puzzle: new minimization algorithms, heuristic scans, parameter kicking and an extended set of physical constraints. It is continuously growing and maturing due to constructive feedback of many individuals. In future, \phoebe aims to broaden its scope on all areas mentioned in this paper: numerical, scientific and technical. We conclude this paper by naming some of the goals \phoebe has yet to achieve.

\begin{enumerate}
\item Full-scale testing. \phoebents's scientific core is now ready for extensive testing on real data. Individual stars as well as large survey databases are ideal testing grounds to hunt down problems and improve those aspects that may now be lagging behind.
\item Scripting. Although the graphical user interface (see Appendix \ref{gui}) is well suited for individual targets, its usability is very limited when the number of EBs is large. We must prepare for the upcoming missions like Gaia, since the shear number of observed EBs will be several orders of magnitude larger than the number of all already solved EBs of today or, for that matter, the number of astronomers in the world to solve them all, one-by-one. It is naive to believe that our procedures are already optimal and applicative to all sorts of EBs that are out there.
\item New physics. Some of the ideas already mentioned in this paper are in their infancy. The SED must evolve into a consistent and reliable data source that enables us to not only confirm or reject the otherwise obtained solution, but to extract parameters from the spectra themselves. Once the SEDs are fully integrated in solution seeking, LD coefficients will have become obsolete, for intensities will then be computable from spectra and the LD effect will come out naturally. Individual components may be intrinsically variable and common types of variabilities may easily be recovered from the model (see \cite{dallaporta2002} for an example of a $\delta$-Sct companion).
\item New numerical algorithms. By the ever-growing computer power, better and more powerful numerical algorithms are surfacing. Two very promising candidates are already in testing: Adaptive simulated annealing \citep{ingber1996} and Powell's direction set method \citep{acton1990}. Both are based only on function evaluations, not numerical derivatives.
\item New technical enhancements. With continuous help and support from users sharing their opinions and suggestions on \phoebe discussion mailing lists, we are able to form a wish-list and implement most needed features. Custom user-supplied passband transmission functions must be supported to enable the data obtained by any instrument and any filter-set to be processed.
\end{enumerate}

Since \phoebe is free (released under the GNU GPL -- see Appendix), everyone with enthusiasm and interest may join in on this project! \phoebe will keep improving.

\acknowledgements
The authors would like to express their utmost gratitude to Robert E.~Wilson, for spending numerous hours commenting and criticising the manuscript and for his continuous encouragement. We are also indebted to the referee of the paper, for making valuable suggestions that significantly improved the paper's layout and clarity. Fruitful discussions with Dirk Terrell, Michael Bauer, Walter Van Hamme, Michael Sallman and Ulisse Munari throughout \phoebe development are very much appreciated. Our thanks go to all \phoebe users out there, supporting our work with constructive feedback. We would also like to thank Phillip J.~Flower for his swift reply on updated coefficients given in Table \ref{flower_cmd}. This work is supported by the Slovenian Ministry for High Education, Science and Technology.

\appendix

\section{Appendix: technical information}

\phoebe is released under the GNU General Public License (GPL) and is freely available for download from \verb|http://phoebe.fiz.uni-lj.si|. The package comes with thorough documentation: Reference manual, Tutorial and Application Programming Interface. For convenience, three discussion mailing lists are available to help users communicate and share opinions, ideas and enhancements to \phoebents. The freedom of GPL enables anyone with interest to join in on future development.

\subsection{The back-end: \phoebe scripter} \label{scripter}

In its core, \phoebe is a \emph{scripting language}. This means that the user communicates with the program interactively by passing particular statements to perform particular actions. 

\phoebe language is based on formal, context-free LALR(1) grammar. This means that strict and consistent grammar rules of scanning, parsing and evaluating user input are imposed to achieve full support for arithmetics, nested loops, conditionals and function definitions (see \cite{aho1986} for specific properties of LALR(1) grammar). It is written in {\tt ANSI C}, which makes it portable to virtually any platform regardless of the operating system used.

\phoebe scripter consists of three layers. The lower-most layer is the {\sf WD} code, the layer above is {\tt PHOEBE}'s extension layer and the topmost layer is the interpreter. The underlying {\sf WD} code is unchanged, which makes adaption to any future {\sf WD} versions trivial. The extension layer contains scientific add-ons that enhance basic {\sf WD} applicativity. Finally, the interpreter's purpose is to communicate with the user. Its plug-in awareness allows miscellaneous technical enhancements to be easily incorporated -- the graphical user interface (GUI), main-sequence calculators etc.

\subsection{The front-end: \phoebe graphical user interface} \label{gui}

All novel \phoebe features discussed in the main part of the paper are implemented in \phoebents's back-end engine, the scripter. Although it is gratifying to achieve advancements in scientific and numerical approaches, technical details that make scientist's life easier are all-too-often overlooked. Based on current \phoebe users' feedback, the most prominent enhancement \phoebe brings into the field is neither numerical nor scientific, it is technical: a graphical user interface (GUI). No longer is it necessary to spend hours or even days learning the technicalities of a particular code; \phoebe features a full-fledged, flexible and heavily structured GUI that brings the ease of clicking, observing and monitoring the process of solution seeking to the user.

\paragraph{The GUI.}

Any implementation of a front-end is inherently system-dependent, and so is \phoebents's interface. The GUI is designed to run under any Linux (or other Linux-compatible) operating system. It should be noted nevertheless that the GUI is merely a plug-in to \phoebe scripter, so when a need for a different GUI on a different operating system arises, it is only a matter of building a front-end -- the back-end will remain the same, portable to all ANSI-C compliant architectures.

\begin{figure}
\begin{center}
\includegraphics[width=12cm]{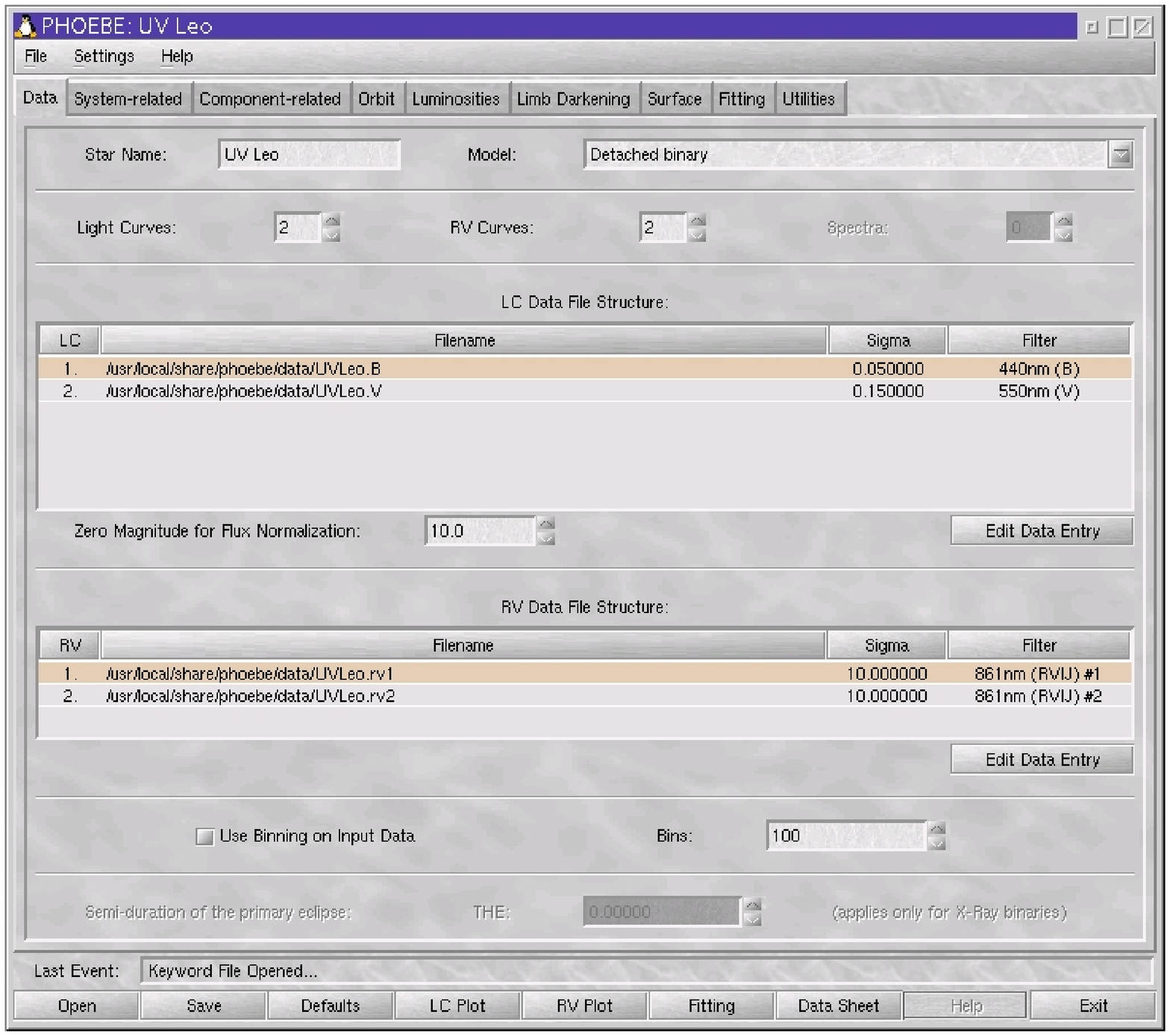}
\caption{A snapshot of \phoebe graphical user interface in action.} \label{phoebegui}
\end{center}
\end{figure}

\phoebents's GUI is written with \verb|GTK+| graphical library, a free standard component of virtually any Linux of today. It consists of the main screen, the snapshot of which is depicted in Fig.~\ref{phoebegui}. The main window is used for basic user interaction - changing parameter values, obtaining statistics on observations, plotting star figures etc. From the button menu on the bottom of the main window users may open auxiliary windows. They are used to plot photometric light curves, RV curves, to initiate the fit or to write \phoebe scripts. The interface is consistent with the rest of the operating system, so users with elementary Linux experience should have \phoebe up and running in no time.

\paragraph{Plotting observations and solutions, star figures.}

Plotting light curves and RV curves throughout the fitting process proves to be extremely useful for consistency checking. \phoebe supports data binning, overplotting, phase aliasing and plotting $O-C$ residuals. In addition, \phoebe plots star figures as they appear on the sky (see the lower right panel of Fig.~\ref{synthetic_lc_rv_starplot} for an example) in any given phase. It is also capable of enhancing the location of spots, making animated cartoons or e.g.~depicting apsidal motion of eccentric binaries.

\paragraph{LD interpolation functions.}

Another very important aspect of modeling EBs that was the sole responsibility of {\sf WD} users are the limb darkening (LD) coefficients. Rather than trusting {\sf WD} to retrieve the values of the coefficients by numerical fits, it is much safer to use precomputed LD tables, e.g.~\citet{vanhamme1993}. \phoebe can retrieve correct values of these coefficients from external LD tables at each iteration step. This speeds up convergence, improves consistency of the derived solution and avoids the need for manual adjustment during the iteration process.

\paragraph{Measuring parameter correlation, handling degeneracies.}

Throughout the paper we have discussed the dark world of parameter correlations and degeneracies. It is crucial for any user to be aware of these problems. An excellent review on what one may expect from light curve modeling is given by \citet{wilson1994}, which is both informative and entertaining, arming the user against the caveats that await him. This is why in addition to {\sf WD}'s correlation matrix \citep{wilson2003} \phoebe plots parameter histograms and convergence tracers during the fit. If we are able to see and inspect the solution in each iteration step and, perhaps more importantly, if we are able to play with the solution as we see fit, we will have more fun with it and more consistent physics is bound to emerge.


\begin{thebibliography}{99}

\bibitem[Acton(1990)]{acton1990} Acton, F.~S.~1990, ISBN 0-883-85450-3

\bibitem[Aho et al.(1986)]{aho1986} Aho, A.~V., Sethi, R.~\& Ullman, J.~D.~1986, ISBN 0-201-10088-6 

\bibitem[Bryja \& Sandtorf(1999)]{bryja1999} Bryja, C.~\& Sandtorf, J.~R.~1999, AAS Meeting 194

\bibitem[Caldwell et al.(1993)]{caldwell1993} Caldwell, J.~A.~R.~et al.~1993, SAAO Circulars, no.~15

\bibitem[Cardelli et al.(1989)]{cardelli1989} Cardelli, J.~A., Clayton, G.~C.~and Mathis, J.~S.~1998, \apj, 345, 245

\bibitem[Claret(2000)]{claret2000} Claret, A.~2000, \aap, 363, 1081

\bibitem[Dallaporta et al.(2002)]{dallaporta2002} Dallaporta, S., Tomov, T., Zwitter, T., Munari, U.~2002, IBVS, 5312

\bibitem[Flower(1996)]{flower1996} Flower, P.~J.~1996, \apj, 469, 355

\bibitem[Forbes et al.(1996)]{forbes1996} Forbes, M.~C., Dodd, R.~J., Sullivan, D.~J.~1996, Balt.~A., 5, 281

\bibitem[Galassi et al.(2003)]{galassi2003} Galassi, M.~et al.~2004, Network Theory Ltd., ISBN 0-954-16173-4

\bibitem[Henden \& Honeycutt(1997)]{henden1997} Henden, A.~\& Honeycutt, R.~K.~1997, \pasp, 109, 441

\bibitem[Henden \& Munari(2000, 2001)]{henden2000} Henden, A.~\& Munari, U.~2000, \aap~Supl.~Ser., 143, 343

\bibitem[Henden \& Munari(2001)]{henden2001} Henden, A.~\& Munari, U.~2001, \aap, 372, 145

\bibitem[Ingber(1996)]{ingber1996} Ingber, L.~1996, Control \& Cybernetics, 25, 33

\bibitem[Kallrath \& Linnell(1987)]{kallrath1987} Kallrath, J.~\& Linnell, A.~P.~1987, \apj, 313, 346

\bibitem[Kallrath et al.(1998)]{kallrath1998} Kallrath, J., Milone, E.~F., Terrell, D.~\& Young, A.~T.~1998, \apj, 508, 308

\bibitem[Kallrath \& Milone(1999)]{kallrath1999} Kallrath, J.~\& Milone, E.~F.~1999, ISBN: 0-387-98622-7 

\bibitem[Kurucz(1998)]{kurucz1998} Kurucz, R.~L.~1998, IAUS, 189, 217

\bibitem[Landolt(1992)]{landolt1992} Landolt, A.~U.~1992, \aj, 104, 340

\bibitem[Lang et al.(1992)]{lang1992} Lang, K.~R.~et al.~1992 (2nd edition), ISBN 0-387-55040-2

\bibitem[Malkov(2003)]{malkov2003} Malkov, O.~Y.~2003, \aap, 402, 1055

\bibitem[Marrese et al.(2004)]{marrese2004} Marrese, P.~M., Munari, U., Sordo, R., Dallaporta, S., Siviero, A., Zwitter, T.~2005, \aap, accepted. astro-ph/0411723

\bibitem[Milone et al.(1992)]{milone1992} Milone, E.~F., Stagg, C.~R., Kurucz, R.~L.~1992, \apjs, 79, 123

\bibitem[Moro \& Munari(2000)]{moro2000} Moro, D.~\& Munari, U.~2000, A\&A Supl.S., 147, 361

\bibitem[Munari et al.(2005)]{munari2004b} Munari, U., Sordo, R., Castelli, F., Zwitter, T.~2005, \aap, submitted

\bibitem[Murphy \& Meiksin(2004)]{murphy2004} Murphy, T.~\& Meiksin, A.~2004, \mnras, 351, 1430

\bibitem[Nelder \& Mead(1965)]{nelder1965} Nelder, J.~A.~\& Mead, R.~1965, Comput. J. 7, 308

\bibitem[Perryman et al.(2001)]{perryman2001} Perryman, M.~A.~C., de Boer, K.~S., Gilmore,~G., H\o g, E., Lattanzi, M.~G., Lindegren,~L., Luri, X., Mignard, F., Pace, O., de Zeeuw, P.~T.~2001, \aap, 369, 339

\bibitem[Pols et al.(1995)]{pols1995} Pols, O.~R., Tout, C.~A.; Eggleton, P.~P.,
Han, Z.~1995, MNRAS, 274, 964

\bibitem[Press et al.(1992)]{nr1992} Press, W.~H., Teukolsky, S.~A., Vetterling, W.~T., Flannery, B.~P.~1992, ISBN~0-521-43108-5

\bibitem[Pr\v sa(2003)]{prsa2003} Pr\v sa, A.~2003, ASP\,\,\,Conf. Ser., 298, 457

\bibitem[Pr\v sa \& Zwitter(2004)]{prsa2004} Pr\v sa, A.~\& Zwitter, T.~2004, \apss, astro-ph/0405314

\bibitem[Pr\v sa \& Zwitter(2005)]{prsa2005} Pr\v sa, A.~\& Zwitter, T.~2005, ASP\,\,\,Conf. Ser., astro-ph/0411264

\bibitem[Sbordone et al.(2004)]{sbordone2004} Sbordone, L, Bonifacio, P., Castelli, F., Kurucz, R.~2004, astro-ph/0406268

\bibitem[Schlegel et al.(1998)]{schlegel1998} Schlegel, D., Finkbeiner, D.~P., Davis, M.~1998, \apj, 500, 525

\bibitem[Siviero et al.(2004)]{siviero2004} Siviero, A., Munari, U., Sordo, R., Dallaporta, S., Marrese, P.~M., Zwitter, T., Milone, E.~F.~2004, \aap, 417, 1083

\bibitem[Terrell et al.(2003)]{terrell2003} Terrell, D., Munari, U., Zwitter, T., Nelson, R.~H.~2003, \aj, 126, 2988

\bibitem[Van Hamme(1993)]{vanhamme1993} Van Hamme, W.~1993, \aj, 106, 2096

\bibitem[Van Hamme \& Wilson(2003)]{vanhamme2003} Van Hamme, W.~\& Wilson, R.~E.~2003, ASP\,\,\,Conf. Ser., 298, 323

\bibitem[Wilson \& Biermann(1976)]{wilson1976a} Wilson, R.~E., Biermann, P.~1976, \aap, 48, 349

\bibitem[Wilson \& Sofia(1976)]{wilson1976} Wilson, R.~E., Sofia, S.~1976, \apj, 203, 182

\bibitem[Wilson(1979)]{wilson1979} Wilson, R.~E.~1979, \apj, 234, 1054

\bibitem[Wilson(1990)]{wilson1990} Wilson, R.~E.~1990, \aj, 356, 613

\bibitem[Wilson(1993)]{wilson1993} Wilson, R.~E.~1993, NFBS, 91

\bibitem[Wilson(1994)]{wilson1994} Wilson, R.~E.~1994, IAPPP, 55, 1

\bibitem[Wilson \& Devinney(1971)]{wilson1971} Wilson, R.~E.~\& Devinney, E.~J.~1971, \apj, 166, 605

\bibitem[Wilson \& Van Hamme(2003)]{wilson2003} Wilson, R.~E.~\& Van Hamme, W.~2003, booklet accompanying WD2003 code

\bibitem[Wyithe \& Wilson(2001)]{wilson2001} Wyithe, J.~S.~B.~\& Wilson, R.~E.~2001, \apj, 559, 260

\bibitem[Zwitter et al.(2004)]{zwitter2004} Zwitter, T., Castelli, F. \& Munari, U.~2004, \aap, 417, 1055

\end{thebibliography}
\end{document}